\documentclass[11pt,a4paper]{article}
\usepackage{jheppub}
\usepackage{amsmath,amsfonts,amssymb}
\usepackage[T1]{fontenc}
\usepackage[utf8]{inputenc}
\usepackage[british]{babel}
\usepackage{lmodern}
\usepackage{tikz}
\usepackage{slashed}

\newcommand{\eq}[1]{\eqref{eq:#1}}
\newcommand{\fig}[1]{Fig.~\ref{fig:#1}}
\newcommand{\tab}[1]{Tab.~\ref{tab:#1}}
\newcommand{\perm}[1]{\text{#1 perm.}}
\newcommand{\tr}{\mathrm{tr}}
\newcommand{\mf}{\mathrm{m}}
\newcommand{\cell}[2]{
  \begin{tabular}{#1} #2 \end{tabular}
}
\newcommand{\ccy}[1]{\mathfrak{y}^{(3)}_{#1}\,}
\newcommand{\ccg}[1]{\mathfrak{g}^{(4)}_{#1}\,}
\newcommand{\openparams}{12}

\preprint{DO-TH 21/02}
\toccontinuoustrue
\begin{document}
\author[a,b]{Tom Steudtner}
\affiliation[a]{
Fakultät für Physik, TU Dortmund, Otto-Hahn-Str. 4, D-44221 Dortmund, Germany
}
\affiliation[b]{
Department of Physics and Astronomy, University of Sussex, Brighton, BN19QH, United Kingdom
}
\emailAdd{tom2.steudtner@tu-dortmund.de}
\title{Towards general scalar-Yukawa renormalisation group equations at three-loop order}
\abstract{ For arbitrary four-dimensional quantum field theories with scalars and fermions, renormalisation group equations in the $\overline{\text{MS}}$ scheme are investigated at three-loop order in perturbation theory. Collecting literature results, general expressions are obtained for field anomalous dimensions, Yukawa interactions, as well as fermion masses. The renormalisation group evolution of scalar quartic, cubic and mass terms is determined up to a few unknown coefficients.
The combined results are applied to compute the renormalisation group evolution of the gaugeless Litim-Sannino model.
}
\maketitle
\newpage
\section{Introduction}
The determination of renormalisation group equations (RGEs) has remained an active field, driven by applications in precision calculations, critical phenomena, model building, grand unification, UV completion and many more. While a plethora of perturbative results exist for theories of special interest, there is also a solid  underpinning of two-loop RGEs for all renormalisable quantum field theories (QFTs) \cite{Machacek:1983tz,Machacek:1983fi,Machacek:1984zw,Luo:2002ti,Schienbein:2018fsw, Sperling:2013eva, Sperling:2013xqa, Sartore:2020pkk}, as well as three-loop order expressions for the gauge sector \cite{Pickering:2001aq, Mihaila:2012pz, Mihaila:2014caa, Poole:2019kcm}. 
This is possible due to considering a template action \cite{Machacek:1983tz,Machacek:1983fi,Machacek:1984zw,Poole:2019kcm,Sartore:2020pkk} of generalised couplings and fields, in which any renormalisable QFT can be embedded. For such a model, loop integrations and spinor summations can be performed explicitly when computing RGEs, while retaining contractions of the generalised couplings.
Thus, results are universally reusable with the reduced complication of inserting embedded couplings into template expressions. In fact, this step can be automatised by existing software tools, such as \cite{Staub:2013tta,Sartore:2020gou,Litim:2020jvl,Thomsen:2021ncy}.
An extension of template framework to the next loop order represents a monumental step which would benefit for all research areas making use of QFT. However, due to the maximal generality of the template action, the computations required involve a high number of Feynman diagrams, which are by themselves technically challenging at three-loop level. Moreover, additional complications due to the $\gamma_5$ problem arise \cite{Jegerlehner:2000dz}. 
It is hence appealing to cut corners by computing RGEs directly for QFTs of interest instead, which however fails to deliver a contribution to the bigger picture. 

And yet, such efforts are not in vain. An approach alternative to computing RGEs in the template QFT directly is to formulate a general ansatz for them instead, consisting of all possible combinations of coupling contractions with unknown coefficients. One thus obtains $\beta$- and $\gamma$-functions depending on these open parameters, which can be compared term by term to existing literature results. The advantage is that such results do not need to be as general as the template theory itself. In fact, input models can be relatively simple, as long as each of them contains enough distinctive features and they are available in sufficient numbers. Hence, the problem of conducting a massive and complicated calculation is broken down into several smaller and independent steps, from which a general result can be collected.
 
The approach has previously been employed in, e.g., \cite{Jack:2014pua,Poole:2019txl,Poole:2019kcm} to study Weyl consistency conditions \cite{ Osborn:1991gm,Jack:2013sha}, as well as in \cite{Jack:2018oec,Steudtner:2020tzo} to compute pure scalar RGEs, and \cite{Jack:2013sha} to gain three-loop Yukawa $\beta$-functions for complex scalars and chiral fermions. The former is paving the way to obtain general 4-loop gauge $\beta$-functions -- with SM expressions already available \cite{Davies:2019onf} -- and eventually general 3-loop Yukawa $\beta$-functions. However, three-loop RGEs for the scalar potential are not covered in these efforts. On the other hand, the scalar sector is often not technically natural due to the lack of a protective symmetry, and hence cannot be neglected in RG studies. Therefore, we target scalar quartic and cubic couplings as well as mass terms in this work. Following up on the pure scalar results \cite{Jack:1990eb,Jack:2018oec,Steudtner:2020tzo,Bednyakov:2021clp}, we also include fermionic contributions through Yukawa interactions and masses, while leaving the entirety of the gauge sector for future studies. As our input, we utilise three-loop results in the Two-Higgs-Doublet Model (THDM) \cite{Herren:2017uxn}, the Standard Model (SM) \cite{Chetyrkin:2012rz,Bednyakov:2013eba,Chetyrkin:2013wya,Bednyakov:2013cpa,Bednyakov:2014pia}, Gross-Neveu-Yukawa theories \cite{Zerf:2017zqi,Mihaila:2017ble} as well as supersymmetric RGEs \cite{Jack:1996qq,Parkes:1985hh}.

This paper is organised as follows: Sections \ref{sec:setup} and \ref{sec:gamma5} introduce our notations and conventions. In Sec.~\ref{sec:gamma-s}--\ref{sec:beta-quart}, renormalisation group equations the general scalar-Yukawa models are obtained. As an application,  three-loop RGEs for the Litim-Sannino model in the gaugeless limit are computed for the first time in Sec.~\ref{sec:lisa}, before concluding with Sec.~\ref{sec:dis}.

\section{Setup}\label{sec:setup}
In this work, we consider a general QFT consisting of Weyl fermions $\psi_i$ and real scalar fields $\phi_a$ with the Lagrange density
  \begin{equation}\label{eq:master-template}
  \begin{aligned}
    \mathcal{L} =
     & \  \tfrac{1}{2} \partial^\mu \phi_a \partial_\mu \phi_a + \tfrac{i}{2} \psi^j \tilde{\sigma}^\mu \partial_\mu \psi_j   - \tfrac{1}{2} y^{ajk} \,\phi_a(\psi_j \varepsilon \psi_k)   - \tfrac{1}{24} \lambda_{abcd} \,\phi_a \phi_b \phi_c \phi_d\,\\
     &  - \tfrac{1}{2} \mf^{jk} \,(\psi_j \varepsilon \psi_k)  
-\tfrac{1}{2} m^2_{ab} \, \phi_a \phi_b - \tfrac{1}{6} h_{abc} \,\phi_a \phi_b \phi_c\,,
  \end{aligned}
  \end{equation}
  featuring Yukawa couplings $y^{aij}$, scalar quartic and cubic interactions $\lambda_{abcd}$, $h_{abc}$ as well as fermion and scalar masses $\mf^{ij}$ and $m^2_{ab}$. Here, $\varepsilon$ denotes the two-dimensional Levi-Civita symbol contracting spinor indices, which are not shown explicitly.
   Each of these interactions are symmetric under permutation of their fermionic $(i,j,k,l,...)$ and scalar indices $(a,b,c,d,...)$, respectively. Moreover, fermion indices in this convention run over Weyl components as well as their conjugates. This accounts for the additional factor $\tfrac12$ and $\tilde{\sigma}^\mu = \sigma^\mu$ or $\overline{\sigma}^\mu$ in the kinetic term, as well the missing complex conjugation of the fermion mass and Yukawa interactions.
  Fields and couplings
  \begin{equation}\label{eq:raise-lower}
     \psi^i = \psi_i^*\,, \quad y^{ajk} \,\phi_a(\psi_j \varepsilon \psi_k) = y^{a}_{jk} \,\phi_a(\psi^j \varepsilon \psi^k) \quad \text{ and } \quad  \mf^{jk} \,(\psi_j \varepsilon \psi_k) = \mf_{jk} \,(\psi^j \varepsilon \psi^k) 
  \end{equation}
  with raised fermionic indices denote the conjugation of Weyl components. This distinction is convenient to ensure that index contractions as in $\mf^{ij} \mf_{jk}$ are due to the propagation of a fermion $\langle\psi_j \,\psi^j\rangle$. Since $\psi_i$ already contains all Weyl fermions as well as their conjugates, $\psi^i$ has the same degrees of freedom. Raising and lowering of fermionic indices merely represents a pairwise permutation of components in $\psi$, such that \eq{raise-lower} holds.
  Note that our notation is very close to the one suggested in \cite{Poole:2019kcm}, where each component of a fermion $\Psi$ contains both a Weyl spinor and its conjugate. Raising and lowering of fermionic indices in \cite{Poole:2019kcm} is made explicit by multiplication with the Pauli matrices $\sigma_1$.
  
   As the scalars are expressed in terms of real components, the raised or lowered position of their indices has no significance.\footnote{For instance, quartic couplings $\lambda_{abcd}$ appear with lowered, while Yukawas $y^a$ with raised scalar indices in this work, which has no other significance but being a style preferred by the author.}  
    In the following, we suppress fermionic indices for convenience, where it is understood that, e.g., $y^a y^b \mf \, y^c  = y^{aij}\, y^{b}_{jk} \,\mf^{kl} \, y^{c}_{lr}$, and the additional abbreviation $y^a y^b y^c y^d ... = y^{abcd...}$ is utilised.
    
Connecting to the notation of e.g. \cite{Machacek:1983tz,Machacek:1983fi,Machacek:1984zw}, where Weyl fermions $\Psi_i$ and their conjugates $\Psi_i^*$ are treated separately
\begin{equation}
\begin{aligned}
  \mathcal{L} = &\phantom{-} i \Psi_j^\dagger \overline{\sigma}^\mu \partial_\mu \Psi_j - \tfrac{1}{2} Y^a_{jk} \,\phi_a(\Psi_j \varepsilon \Psi_k) - \tfrac{1}{2} {Y^\dagger}^a_{jk} \,\phi_a(\Psi_j^* \varepsilon \Psi_k^*) \\
  & - \tfrac{1}{2} \mf_{jk} \,(\Psi_j \varepsilon \Psi_k) - \tfrac{1}{2} \mf_{jk}^\dagger \,(\Psi_j^* \varepsilon \Psi_k^*) + \dots, 
\end{aligned}
\end{equation}
the relation 
  \begin{equation}
    \tr(y^a y^b y^c y^d ...) = \tr({Y}^a Y^{\dagger b} Y^c Y^{\dagger d } ...) + \tr(Y^{\dagger a} {Y}^b Y^{\dagger c} Y^d ...)
  \end{equation}
  holds, while untraced products are simply expressed via $Y^a Y^{\dagger b} Y^c ... = y^{abc...}$. 
  
  Through the course of this paper, we determine the RG evolution for general QFTs without gauge interactions at three-loop in perturbation theory. At this loop order, results depend on the regularisation and renormalisation procedure, for which we employ dimensional regularisation (DREG) \cite{Bollini:1972bi,Bollini:1972ui} and the $\overline{\text{MS}}$ scheme \cite{tHooft:1973mfk,Bardeen:1978yd}. Due to this choice, there is also an inherent ambiguity in the treatment of $\gamma_5$, which is discussed in the next section.
  As all gaugeless and renormalisable QFTs can be mapped onto our template \eq{master-template}, we aim to determine $\beta$- and $\gamma$-functions for its generalised couplings and fields. 
Overall, we will proceed to formulate our ansätze for three-loop RGEs in terms of contractions of the dimensionless couplings $\lambda_{abcd}$ and $y^a$ only, and employ the dummy field trick \cite{Martin:1993zk,Luo:2002ti,Schienbein:2018fsw} to extract $\beta$-functions for $h_{abc}$, $m^2_{ab}$ and $\mf^{ij}$. To facilitate the evaluation, a modified version of \texttt{ARGES} \cite{Litim:2020jvl} is employed. 

  \section{Handling of \texorpdfstring{$\mathbf{\gamma_5}$}{gamma\_5}}\label{sec:gamma5}  
  The DREG procedure prescribes the evaluation of loop integrals in $d = 4 - 2 \epsilon$ instead of four dimensions. However, this is incompatible with the spinor algebra of Dirac matrices $\gamma^\mu$ \cite{tHooft:1972tcz}, as the definition
  \begin{equation}
    \gamma_5 = \tfrac{i}{24}\, \epsilon_{\mu\nu\rho\sigma} \gamma^\mu \gamma^\nu \gamma^\rho \gamma^\sigma  \qquad \text{ with } \qquad \epsilon_{0123} = -\epsilon^{0123} = 1
  \end{equation}
 being the Levi-Civita symbol, is closely tied to four dimensions. Different approaches to reconcile the $\gamma_5$ problem -- see \cite{Jegerlehner:2000dz} for a review and \cite{Belusca-Maito:2020ala} for a recent list of works -- in general lead to ambiguous computational results; this includes renormalisation group equations, see e.g. \cite{Bednyakov:2015ooa,Zoller:2015tha}.
 The source of these ambiguities is the occurrence of the expression
   \begin{equation}\label{eq:gamma5-term}
     \tr \left[\gamma^\mu \gamma^\nu \gamma^\rho \gamma^\sigma \gamma_5 \right] = \begin{cases}
    \ 4i\,\epsilon^{\mu\nu\rho\sigma} & \text{ for } d = 4,\\
    \ 0 & \text{ na\"ively in } d \text{ dimensions with } \{\gamma^\mu, \gamma_5\}=0
     \end{cases}
   \end{equation}
in loop corrections.
 In \cite{Chetyrkin:2012rz,Bednyakov:2012en}, a \textit{semi-na\"ive} scheme has been chosen that renders the ambiguity evanescent $(\propto d\!-\!4)$ and hence drop out when computing three-loop Yukawa $\beta$-functions. This scheme has also been employed in \cite{Bednyakov:2014pia,Zerf:2017zqi,Mihaila:2017ble,Herren:2017uxn}, which serve as input in this work. At four-loop order, the approach is not sufficient, but contributions affected by the $\gamma_5$ problem in \cite{Poole:2019txl,Poole:2019kcm,Davies:2019onf} have been fixed using Weyl consistency conditions. 
 
In our setup, tensor structures that explicitly distinguish the chiralities of the included fermion lines have to be considered in the ansatz in order to account for possible $\gamma_5$ corrections. This could be achieved by inserting the quantity $\chi^i_{\phantom{i}j} = \pm\, \delta^i_{\phantom{i}j}$ on fermion lines, providing opposite signs for left- and right-chiral fermions. Obviously, this is a direct adaptation of $\gamma_5$ into the language of Weyl components.
However, due to the absence of gauge interactions, there are actually no $\gamma_5$ ambiguities in this work. 
Following the argumentation in \cite{Chetyrkin:2012rz,Bednyakov:2012en,Poole:2019txl,Poole:2019kcm}, a non-vanishing term \eq{gamma5-term} only occurs if there is a combination of four independent momenta and/or Lorentz indices. All RGEs are independent of external momenta, which can be set to zero in the final result.
As gauge fields are not present in our work, four independent loop momenta are required to generate an ambiguity from \eq{gamma5-term}. This, however, cannot occur at three-loop order as investigated here, but would become relevant at four-loop. In fact, all ambiguous three-loop terms in the general Yukawa $\beta$-function, as well as scalar and fermion field anomalous dimensions, have been identified in \cite{Poole:2019txl,Poole:2019kcm} to contain gauge interactions. Consequently, the $\gamma_5$ issue can be neglected both in our ansatz as well as in the input data we use, as the gauge sector will be projected out.

\section{Scalar Anomalous Dimension}\label{sec:gamma-s}
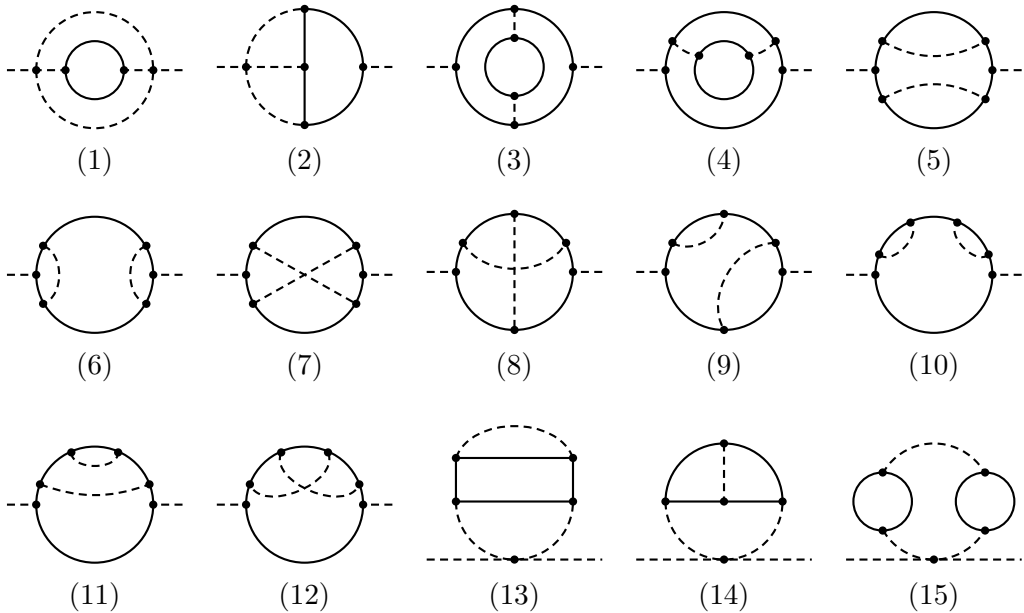
\begin{figure}[ht]
  \centering
  \begin{tabular}{ccccc}
    \begin{tikzpicture}
    \draw[black, thick, densely dashed] (1.em, 0.em) -- (3em, 0em);
    \draw[black, thick, densely dashed] (7em, 0.em) -- (5em, 0em);
    \draw[black, thick, densely dashed] (2em, 0.em) arc (180:0:2.em);
    \draw[black, thick, densely dashed] (2em, 0.em) arc (180:360:2.em);
    \draw[black, thick] (4em, 0.em) circle (1em);
    \node at (2em, 0)[circle,fill,inner sep=.1em]{};
    \node at (6em, 0)[circle,fill,inner sep=.1em]{};
    \node at (3em, 0)[circle,fill,inner sep=.1em]{};
    \node at (5em, 0)[circle,fill,inner sep=.1em]{};
    \end{tikzpicture}
    &
    \begin{tikzpicture}
    \draw[black, thick, densely dashed] (1.em, 0.em) -- (4em, 0em);
    \draw[black, thick, densely dashed] (7em, 0.em) -- (6em, 0em);
    \draw[black, thick, densely dashed] (2em, 0.em) arc (180:90:2.em);
    \draw[black, thick, densely dashed] (2em, 0.em) arc (180:270:2.em);
    \draw[black, thick] (6em, 0.em) arc (0:90:2.em);
    \draw[black, thick] (6em, 0.em) arc (0:-90:2.em) -- (4em, 2.em);
    \node at (2em, 0)[circle,fill,inner sep=.1em]{};
    \node at (6em, 0)[circle,fill,inner sep=.1em]{};
    \node at (4em, 0)[circle,fill,inner sep=.1em]{};
    \node at (4em, 2em)[circle,fill,inner sep=.1em]{};
    \node at (4em, -2em)[circle,fill,inner sep=.1em]{};
    \end{tikzpicture}
    &
    \begin{tikzpicture}
    \draw[black, thick, densely dashed] (1.em, 0.em) -- (2em, 0.em);
    \draw[black, thick, densely dashed] (7.em, 0.em) -- (6em, 0.em);
    \draw[black, thick] (4em, 0em) circle (2em);
    \draw[black, thick] (4em, 0em) circle (1em);
    \draw[black, thick, densely dashed] (4em, 2.em) -- (4em, 1.em);
    \draw[black, thick, densely dashed] (4em, -2.em) -- (4em, -1.em);
    \node at (2em, 0.em)[circle,fill,inner sep=.1em]{};
    \node at (6em, 0.em)[circle,fill,inner sep=.1em]{};
    \node at (4em, 2.em)[circle,fill,inner sep=.1em]{};
    \node at (4em, -2.em)[circle,fill,inner sep=.1em]{};
    \node at (4em, 1.em)[circle,fill,inner sep=.1em]{};
    \node at (4em, -1.em)[circle,fill,inner sep=.1em]{};
    \end{tikzpicture}
     &
     \begin{tikzpicture}
    \draw[black, thick, densely dashed] (1.em, 0.em) -- (2em, 0.em);
    \draw[black, thick, densely dashed] (7.em, 0.em) -- (6em, 0.em);
    \draw[black, thick] (4em, 0em) circle (2em);
    \draw[black, thick] (4em, 0em) circle (1em);
    \draw[black, thick, densely dashed] (2.236em, 1em) to [bend right=20] (3.15em, .5em);
    \draw[black, thick, densely dashed] (5.764em, 1em) to [bend left=20]  (4.85em, .5em);
    \node at (2em, 0.em)[circle,fill,inner sep=.1em]{};
    \node at (6em, 0.em)[circle,fill,inner sep=.1em]{};
    \node at (2.236em, 1em)[circle,fill,inner sep=.1em]{};
    \node at (5.764em, 1em)[circle,fill,inner sep=.1em]{};
    \node at (3.15em, .5em)[circle,fill,inner sep=.1em]{};
    \node at (4.85em, .5em)[circle,fill,inner sep=.1em]{};
    \end{tikzpicture}
     &
    \begin{tikzpicture}
    \draw[black, thick, densely dashed] (1.em, 0.em) -- (2em, 0.em);
    \draw[black, thick, densely dashed] (7.em, 0.em) -- (6em, 0.em);
    \draw[black, thick] (4em, 0em) circle (2em);
    \draw[black, thick, densely dashed] (2.236em, 1em) to [bend right] (5.764em, 1em);
    \draw[black, thick, densely dashed] (2.236em, -1em) to [bend left] (5.764em, -1em);
    \node at (2em, 0.em)[circle,fill,inner sep=.1em]{};
    \node at (6em, 0.em)[circle,fill,inner sep=.1em]{};
    \node at (2.236em, 1em)[circle,fill,inner sep=.1em]{};
    \node at (2.236em, -1em)[circle,fill,inner sep=.1em]{};
    \node at (5.764em, 1em)[circle,fill,inner sep=.1em]{};
    \node at (5.764em, -1em)[circle,fill,inner sep=.1em]{};
    \end{tikzpicture}
    \\ (1) & (2) & (3) & (4) & (5)\\[1em]
    \begin{tikzpicture}
    \draw[black, thick, densely dashed] (1.em, 0.em) -- (2em, 0.em);
    \draw[black, thick, densely dashed] (7.em, 0.em) -- (6em, 0.em);
    \draw[black, thick] (4em, 0em) circle (2em);
    \draw[black, thick, densely dashed] (2.236em, 1em) to [bend left=70] (2.236em, -1em);
    \draw[black, thick, densely dashed] (5.764em, 1em) to [bend right=70]  (5.764em, -1em);
    \node at (2em, 0.em)[circle,fill,inner sep=.1em]{};
    \node at (6em, 0.em)[circle,fill,inner sep=.1em]{};
    \node at (2.236em, 1em)[circle,fill,inner sep=.1em]{};
    \node at (2.236em, -1em)[circle,fill,inner sep=.1em]{};
    \node at (5.764em, 1em)[circle,fill,inner sep=.1em]{};
    \node at (5.764em, -1em)[circle,fill,inner sep=.1em]{};
    \end{tikzpicture}
    &
    \begin{tikzpicture}
    \draw[black, thick, densely dashed] (1.em, 0.em) -- (2em, 0.em);
    \draw[black, thick, densely dashed] (7.em, 0.em) -- (6em, 0.em);
    \draw[black, thick] (4em, 0em) circle (2em);
    \draw[black, thick, densely dashed] (2.236em, 1em) to (5.764em, -1em);
    \draw[black, thick, densely dashed] (2.236em, -1em) to (5.764em, 1em);
    \node at (2em, 0.em)[circle,fill,inner sep=.1em]{};
    \node at (6em, 0.em)[circle,fill,inner sep=.1em]{};
    \node at (2.236em, 1em)[circle,fill,inner sep=.1em]{};
    \node at (2.236em, -1em)[circle,fill,inner sep=.1em]{};
    \node at (5.764em, 1em)[circle,fill,inner sep=.1em]{};
    \node at (5.764em, -1em)[circle,fill,inner sep=.1em]{};
    \end{tikzpicture}
    &
    \begin{tikzpicture}
    \draw[black, thick, densely dashed] (1.em, 0.em) -- (2em, 0.em);
    \draw[black, thick, densely dashed] (7.em, 0.em) -- (6em, 0.em);
    \draw[black, thick] (4em, 0em) circle (2em);
    \draw[black, thick, densely dashed] (2.236em, 1em) to [bend right=60] (5.764em, 1em);
    \draw[black, thick, densely dashed] (4em, 2em) -- (4em, -2em);
    \node at (2em, 0.em)[circle,fill,inner sep=.1em]{};
    \node at (6em, 0.em)[circle,fill,inner sep=.1em]{};
    \node at (2.236em, 1em)[circle,fill,inner sep=.1em]{};
    \node at (5.764em, 1em)[circle,fill,inner sep=.1em]{};
    \node at (4em, 2em)[circle,fill,inner sep=.1em]{};
    \node at (4em, -2em)[circle,fill,inner sep=.1em]{};
    \end{tikzpicture}
    &
    \begin{tikzpicture}
    \draw[black, thick, densely dashed] (1.em, 0.em) -- (2em, 0.em);
    \draw[black, thick, densely dashed] (7.em, 0.em) -- (6em, 0.em);
    \draw[black, thick] (4em, 0em) circle (2em);
    \draw[black, thick, densely dashed] (2.236em, 1em) to [bend right=60] (4em, 2em); 
    \draw[black, thick, densely dashed] (5.764em, 1em) to [bend right=60] (4em, -2em);
    \node at (2em, 0.em)[circle,fill,inner sep=.1em]{};
    \node at (6em, 0.em)[circle,fill,inner sep=.1em]{};
    \node at (2.236em, 1em)[circle,fill,inner sep=.1em]{};
    \node at (5.764em, 1em)[circle,fill,inner sep=.1em]{};
    \node at (4em, 2em)[circle,fill,inner sep=.1em]{};
    \node at (4em, -2em)[circle,fill,inner sep=.1em]{};
    \end{tikzpicture}
    &
    \begin{tikzpicture}
    \draw[black, thick, densely dashed] (1.em, 0.em) -- (2em, 0.em);
    \draw[black, thick, densely dashed] (7.em, 0.em) -- (6em, 0.em);
    \draw[black, thick] (4em, 0em) circle (2em);
    \draw[black, thick, densely dashed] (2.13em, .7em) to [bend right=90] (3.2em, 1.8em);
    \draw[black, thick, densely dashed] (5.87em, .7em) to [bend left=90] (4.8em, 1.8em);
    \node at (2em, 0.em)[circle,fill,inner sep=.1em]{};
    \node at (6em, 0.em)[circle,fill,inner sep=.1em]{};
    \node at (3.2em, 1.8em)[circle,fill,inner sep=.1em]{};
    \node at (2.13em, .7em)[circle,fill,inner sep=.1em]{};
    \node at (4.8em, 1.8em)[circle,fill,inner sep=.1em]{};
    \node at (5.87em, .7em)[circle,fill,inner sep=.1em]{};
    \end{tikzpicture}
    \\ (6) & (7) & (8) & (9) & (10) \\[1em]
    \begin{tikzpicture}
    \draw[black, thick, densely dashed] (1.em, 0.em) -- (2em, 0.em);
    \draw[black, thick, densely dashed] (7.em, 0.em) -- (6em, 0.em);
    \draw[black, thick] (4em, 0em) circle (2em);
    \draw[black, thick, densely dashed] (2.13em, .7em) to [bend right=20]  (5.87em, .7em) ;
    \draw[black, thick, densely dashed] (3.2em, 1.8em) to [bend right=90] (4.8em, 1.8em);
    \node at (2em, 0.em)[circle,fill,inner sep=.1em]{};
    \node at (6em, 0.em)[circle,fill,inner sep=.1em]{};
    \node at (3.2em, 1.8em)[circle,fill,inner sep=.1em]{};
    \node at (2.13em, .7em)[circle,fill,inner sep=.1em]{};
    \node at (4.8em, 1.8em)[circle,fill,inner sep=.1em]{};
    \node at (5.87em, .7em)[circle,fill,inner sep=.1em]{};
    \end{tikzpicture}
    &
    \begin{tikzpicture}
    \draw[black, thick, densely dashed] (1.em, 0.em) -- (2em, 0.em);
    \draw[black, thick, densely dashed] (7.em, 0.em) -- (6em, 0.em);
    \draw[black, thick] (4em, 0em) circle (2em);
    \draw[black, thick, densely dashed] (4.8em, 1.8em) to [bend left=90]  (2.13em, .7em);
    \draw[black, thick, densely dashed] (3.2em, 1.8em) to [bend right=90]  (5.87em, .7em) ;
    \node at (2em, 0.em)[circle,fill,inner sep=.1em]{};
    \node at (6em, 0.em)[circle,fill,inner sep=.1em]{};
    \node at (3.2em, 1.8em)[circle,fill,inner sep=.1em]{};
    \node at (2.13em, .7em)[circle,fill,inner sep=.1em]{};
    \node at (4.8em, 1.8em)[circle,fill,inner sep=.1em]{};
    \node at (5.87em, .7em)[circle,fill,inner sep=.1em]{};
    \end{tikzpicture}
    &
    \begin{tikzpicture}
    \draw[black, thick, densely dashed] (1.em, 0.em) -- (7em, 0.em);
    \draw[black, thick, densely dashed] (2em, 2.em) arc (180:360:2.em);
    \draw[black, thick, densely dashed] (2em, 3.5em) to [bend left=70]  (6em, 3.5em);
    \draw[black, thick] (2em, 2em) -- (2em, 3.5em) --  (6em, 3.5em) -- (6em, 2em) -- cycle ;
    \node at (4em, 0em)[circle,fill,inner sep=.1em]{};
    \node at (2em, 3.5em)[circle,fill,inner sep=.1em]{};
    \node at (6em, 3.5em)[circle,fill,inner sep=.1em]{};
    \node at (2em, 2em)[circle,fill,inner sep=.1em]{};
    \node at (6em, 2em)[circle,fill,inner sep=.1em]{};
    \end{tikzpicture}
    &
    \begin{tikzpicture}
    \draw[black, thick, densely dashed] (1.em, 0.em) -- (7em, 0.em);
    \draw[black, thick, densely dashed] (2em, 2.em) arc (180:360:2.em);
    \draw[black, thick, densely dashed] (4.em, 4.em) -- (4.em, 2.em);
    \draw[black, thick] (2em, 2.em) arc (180:0:2.em) -- cycle;
    \node at (4em, 0em)[circle,fill,inner sep=.1em]{};
    \node at (2em, 2em)[circle,fill,inner sep=.1em]{};
    \node at (6em, 2em)[circle,fill,inner sep=.1em]{};
    \node at (4em, 2em)[circle,fill,inner sep=.1em]{};
    \node at (4em, 4em)[circle,fill,inner sep=.1em]{};
    \end{tikzpicture}
    &
    \begin{tikzpicture}
    \draw[black, thick, densely dashed] (1.em, 0.em) -- (7em, 0.em);
    \draw[black, thick, densely dashed] (4em, 2.em) circle (2em);
    \draw[black, thick, fill=white] (2.25em, 2.em) circle (1em);
    \draw[black, thick, fill=white] (5.75em, 2.em) circle (1em);
    \node at (4em, 0em)[circle,fill,inner sep=.1em]{};
    \node at (5.75em, 1.em)[circle,fill,inner sep=.1em]{};
    \node at (5.75em, 3.em)[circle,fill,inner sep=.1em]{};
    \node at (2.25em, 1.em)[circle,fill,inner sep=.1em]{};
    \node at (2.25em, 3.em)[circle,fill,inner sep=.1em]{};
    \end{tikzpicture}
    \\ (11) & (12) & (13) & (14) & (15) 
  \end{tabular}
  \caption{Fermionic (solid lines) and scalar (dashed lines) diagrams potentially giving rise to three-loop corrections of scalar field anomalous dimensions. One additional diagram providing a purely scalar contribution is not shown.}
  \label{fig:gamma_s}
\end{figure}
The canonical renormalisation procedure of the scalar fields entails the introduction of symmetric field strength renormalisation matrices $Z^\phi_{ab} = Z^\phi_{ba}$ via $\phi_a \mapsto \sqrt{Z^\phi}_{ab}\,\phi_b$ in \eq{master-template}. The running of this quantity under a change of the renormalisation scale $\mu$ is then encoded in the scalar field anomalous dimension
\begin{equation}\label{eq:gamma_s-def}
  \gamma^\phi_{ab} = \frac{\mathrm{d} \sqrt{Z^\phi}_{ac}}{\mathrm{d} \ln \mu} \left(\sqrt{Z^\phi}\right)^{-1}_{cb} = \sum_{n=1}^\infty \frac{\gamma^{\phi,n\ell}_{ab}}{(4\pi)^{2n}}\,.
\end{equation}
The one- and two-loop results $\gamma^{\phi,1\ell}$ and $\gamma^{\phi,2\ell}$ of this quantity are known for arbitrary renormalisable QFTs \cite{Machacek:1983tz,Luo:2002ti}, and the pure scalar (quartic) three-loop contribution to $\gamma^{\phi,3\ell}$  has been found in  \cite{Jack:1990eb,Jack:2018oec,Steudtner:2020tzo,Bednyakov:2021clp}. Extending this progress with an arbitrary Yukawa sector, we find 15 additional contractions, listed in \fig{gamma_s}. Among those are also (2) and (9) which are manifestly asymmetric under exchange of external legs. While $Z^\phi_{ab} = Z^\phi_{ba}$, the square root $\sqrt{Z^\phi}_{ab}$ is only symmetric up to an arbitrary an orthogonal transformation of the scalar field species, which is translated into a potential asymmetry of $\gamma^\phi$ by virtue of \eq{gamma_s-def}. Such rotations in the basis of scalar fields correspond to transformations of running couplings, which drops out in physical observables and are absorbed by the antisymmetric parts of the field anomalous dimensions \cite{Bednyakov:2014pia,Herren:2017uxn,Jack:2016tpp}. 
We use \cite{Herren:2017uxn} to fix all open coefficients, where anomalous dimensions are chosen completely symmetric. This yields
\begin{equation}\label{eq:gamma_s}
  \begin{aligned}
    \gamma_{ab}^{\phi,3\ell} &= - \tfrac{1}{16} \lambda_{acde} \lambda_{defg} \lambda_{bcfg} - \tfrac{5}{32} \lambda_{acde} \lambda_{bcdf} \, \tr (y^{ef}) \\
   &\phantom{=} + \tfrac58\! \left[ \lambda_{acde}\,\tr(y^{bcde}) + \tr(y^{acde})\,\lambda_{bcde}\right]  + \left[\tr(y^{abcd}) + \tfrac9{16}\, \tr(y^{acbd} )\right] \tr(y^{cd}) \\
   &\phantom{=} - \tfrac3{16} \, \tr(y^{abccdd}) + \tfrac5{16}\,\tr(y^{abcddc}) - \tfrac38\, \tr(y^{abcdcd} ) + \tfrac7{16} \left[ \tr(y^{acbcdd}) + \tr(y^{bcacdd})  \right]\\
    &\phantom{=}  - \tfrac34\, \tr(y^{acbdcd}) + \tfrac1{32}\,\tr(y^{accbdd}) + \tfrac74 \,\tr(y^{acdbdc}) + \left[\tfrac32 \zeta_3 -1 \right] \tr(y^{acdbcd}),
  \end{aligned}
\end{equation}
employing the abbreviation $y^{abcd...} = y^a y^b y^c y^d ...$ and $\zeta_3 \approx 1.202$ is Apéry's constant. It is found that the graphs (13)--(15) in \fig{gamma_s} do not contribute to the anomalous dimension, which can be readily understood from the momentum flow.
\section{Fermion Anomalous Dimension}\label{sec:gamma-f}
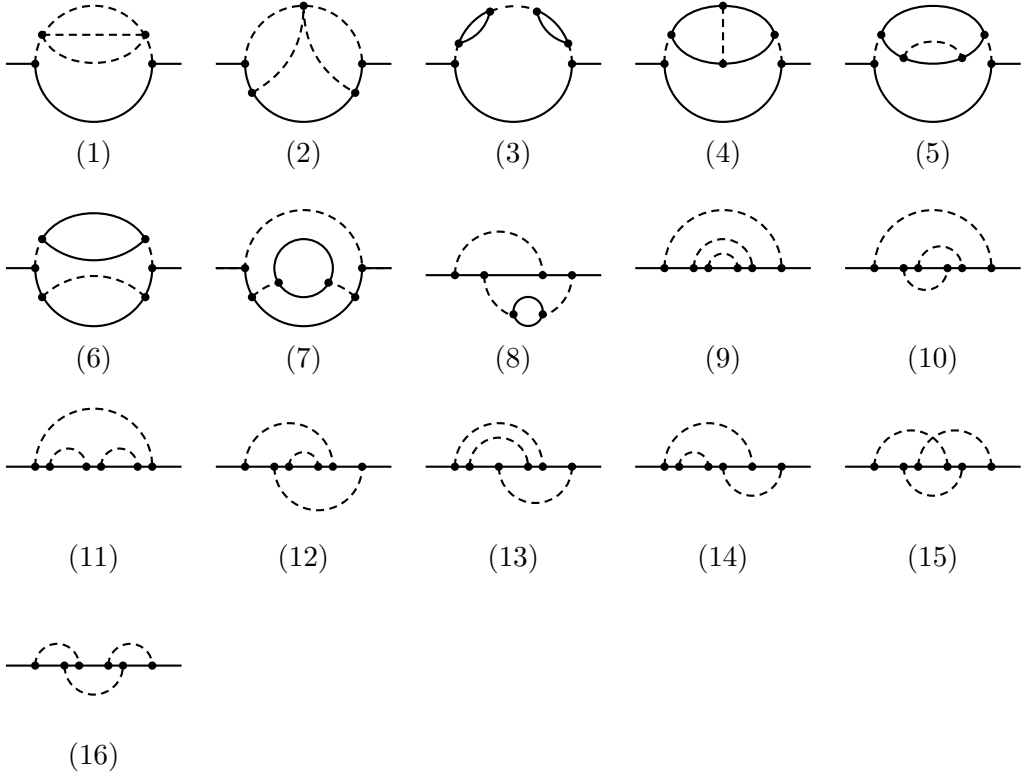
\begin{figure}[ht]
  \centering
  \begin{tabular}{ccccc}
    \begin{tikzpicture}
      \draw[black, thick] (1.em, 0.em) -- (2em, 0.em) arc (180:360: 2em) -- (7.em, 0.em);
      \draw[black, thick]  (6em, 0.em);
      \draw[black, thick, densely dashed] (2.236em, 1em) to [bend right=70] (5.764em, 1em);
       \draw[black, thick, densely dashed] (2.236em, 1em) -- (5.764em, 1em);
       \draw[black, thick, densely dashed] (2em, 0.em) arc (180:0: 2em);
      \node at (2em, 0.em)[circle,fill,inner sep=.1em]{};
      \node at (6em, 0.em)[circle,fill,inner sep=.1em]{};
      \node at (2.236em, 1em)[circle,fill,inner sep=.1em]{};
      \node at (5.764em, 1em)[circle,fill,inner sep=.1em]{};
    \end{tikzpicture}
    &
    \begin{tikzpicture}
    \draw[black, thick] (1.em, 0.em) -- (2em, 0.em) arc (180:360: 2em) -- (7.em, 0.em);
    \draw[black, thick, densely dashed] (2em, 0.em) arc (180:0: 2em);
    \draw[black, thick, densely dashed] (2.236em, -1em) to [bend right] (4em, 2em) to [bend right] (5.764em, -1em);
    \node at (2em, 0.em)[circle,fill,inner sep=.1em]{};
    \node at (6em, 0.em)[circle,fill,inner sep=.1em]{};
    \node at (2.236em, -1em)[circle,fill,inner sep=.1em]{};
    \node at (5.764em, -1em)[circle,fill,inner sep=.1em]{};
    \node at (4em, 2em)[circle,fill,inner sep=.1em]{};
    \end{tikzpicture}
    &
    \begin{tikzpicture}
    \draw[black, thick] (1.em, 0.em) -- (2em, 0.em) arc (180:360: 2em) -- (7.em, 0.em);
    \draw[black, thick, densely dashed] (2em, 0.em) arc (180:160: 2em);
    \draw[black, thick] (2.13em, .7em) arc (160:115: 2em);
    \draw[black, thick, densely dashed] (6em, 0.em) arc (0:20: 2em);
    \draw[black, thick] (5.87em, .7em) arc (20:65: 2em);
    \draw[black, thick,  dashed] (3.2em, 1.8em) arc (115:65: 2em);
    \draw[black, thick] (2.13em, .7em) to [bend right=30] (3.2em, 1.8em);
    \draw[black, thick] (4.8em, 1.8em) to [bend right=30] (5.87em, .7em);
    \node at (2em, 0.em)[circle,fill,inner sep=.1em]{};
    \node at (6em, 0.em)[circle,fill,inner sep=.1em]{};
    \node at (3.2em, 1.8em)[circle,fill,inner sep=.1em]{};
    \node at (2.13em, .7em)[circle,fill,inner sep=.1em]{};
    \node at (4.8em, 1.8em)[circle,fill,inner sep=.1em]{};
    \node at (5.87em, .7em)[circle,fill,inner sep=.1em]{};
    \end{tikzpicture}
    &
    \begin{tikzpicture}
    \draw[black, thick] (1.em, 0.em) -- (2em, 0.em) arc (180:360: 2em) -- (7.em, 0.em);
    \draw[black, thick] (2.236em, 1em) to [bend right=75] (5.764em, 1em);
    \draw[black, thick] (2.236em, 1em) to [bend left=75] (5.764em, 1em);
    \draw[black, thick, densely dashed] (4em, 2em) -- (4em, 0em);
    \draw[black, thick, densely dashed] (2em, 0em) to [bend left=15] (2.236em, 1em);
    \draw[black, thick, densely dashed] (6em, 0em) to [bend right=15] (5.764em, 1em);
    \node at (2em, 0.em)[circle,fill,inner sep=.1em]{};
    \node at (6em, 0.em)[circle,fill,inner sep=.1em]{};
    \node at (2.236em, 1em)[circle,fill,inner sep=.1em]{};
    \node at (5.764em, 1em)[circle,fill,inner sep=.1em]{};
    \node at (4em, 2em)[circle,fill,inner sep=.1em]{};
    \node at (4em, 0)[circle,fill,inner sep=.1em]{};
    \end{tikzpicture}
    &
    \begin{tikzpicture}
    \draw[black, thick] (1.em, 0.em) -- (2em, 0.em) arc (180:360: 2em) -- (7.em, 0.em);
    \draw[black, thick] (2.236em, 1em) to [bend right=75] (5.764em, 1em);
    \draw[black, thick] (2.236em, 1em) to [bend left=75] (5.764em, 1em);
    \draw[black, thick, densely dashed] (2em, 0em) to [bend left=15] (2.236em, 1em);
    \draw[black, thick, densely dashed] (6em, 0em) to [bend right=15] (5.764em, 1em);
    \draw[black, thick, densely dashed] (3em, .2em) to [bend left=75] (5em, .2em);
    \node at (2em, 0.em)[circle,fill,inner sep=.1em]{};
    \node at (6em, 0.em)[circle,fill,inner sep=.1em]{};
    \node at (2.236em, 1em)[circle,fill,inner sep=.1em]{};
    \node at (5.764em, 1em)[circle,fill,inner sep=.1em]{};
    \node at (3em, .2em)[circle,fill,inner sep=.1em]{};
    \node at (5em, .2em)[circle,fill,inner sep=.1em]{};
    \end{tikzpicture}
    \\ (1) & (2) & (3) & (4) & (5) \\[1em]
    \begin{tikzpicture}
    \draw[black, thick] (1.em, 0.em) -- (2em, 0.em) arc (180:360: 2em) -- (7.em, 0.em);
    \draw[black, thick] (2.236em, 1em) to [bend right=45] (5.764em, 1em);
    \draw[black, thick] (2.236em, 1em) to [bend left=60] (5.764em, 1em);
    \draw[black, thick, densely dashed] (2em, 0em) to [bend left=15] (2.236em, 1em);
    \draw[black, thick, densely dashed] (6em, 0em) to [bend right=15] (5.764em, 1em);
    \draw[black, thick, densely dashed] (2.236em, -1em) to [bend left=45] (5.764em, -1em);
    \node at (2em, 0.em)[circle,fill,inner sep=.1em]{};
    \node at (6em, 0.em)[circle,fill,inner sep=.1em]{};
    \node at (2.236em, 1em)[circle,fill,inner sep=.1em]{};
    \node at (5.764em, 1em)[circle,fill,inner sep=.1em]{};
     \node at (2.236em, -1em)[circle,fill,inner sep=.1em]{};
    \node at (5.764em, -1em)[circle,fill,inner sep=.1em]{};
    \end{tikzpicture}
    &
    \begin{tikzpicture}
    \draw[black, thick] (1.em, 0.em) -- (2em, 0.em) arc (180:360: 2em) -- (7.em, 0.em);
    \draw[black, thick, densely dashed] (1.em, 0.em) -- (2em, 0.em) arc (180:0: 2em) -- (7.em, 0.em);
    \draw[black, thick, densely dashed] (2.236em, -1em) to [bend left=15] (3.15em, -.5em);
    \draw[black, thick, densely dashed] (5.764em, -1em) to [bend right=15] (4.85em, -.5em);
    \draw[black, thick] (4.em, 0.em) circle (1.em);
    \node at (2em, 0.em)[circle,fill,inner sep=.1em]{};
    \node at (6em, 0.em)[circle,fill,inner sep=.1em]{};
     \node at (2.236em, -1em)[circle,fill,inner sep=.1em]{};
    \node at (5.764em, -1em)[circle,fill,inner sep=.1em]{};
    \node at (3.15em, -.5em)[circle,fill,inner sep=.1em]{};
    \node at (4.85em, -.5em)[circle,fill,inner sep=.1em]{};
    \end{tikzpicture}
    &
    \begin{tikzpicture}
    \draw[black, thick] (1.em, 0.em) -- (7.em, 0.em);
    \draw[black, thick, densely dashed] (2em, 0.em) arc (180:0: 1.5em);
    \draw[black, thick, densely dashed] (3em, 0.em) arc (180:360: 1.5em);
    \draw[black, thick, fill=white] (4.5em, -1.25em) circle (0.5em);
    \node at (2em, 0.em)[circle,fill,inner sep=.1em]{};
    \node at (6em, 0.em)[circle,fill,inner sep=.1em]{};
    \node at (5em, 0.em)[circle,fill,inner sep=.1em]{};
    \node at (3em, 0.em)[circle,fill,inner sep=.1em]{};
    \node at (5em, -1.35em)[circle,fill,inner sep=.1em]{};
    \node at (4em, -1.35em)[circle,fill,inner sep=.1em]{};
    \end{tikzpicture}
    &
    \begin{tikzpicture}
    \draw[black, thick] (1.em, 0.em) -- (7.em, 0.em);
    \draw[black, thick, densely dashed] (2em, 0.em) arc (180:0: 2em);
    \draw[black, thick, densely dashed] (3em, 0.em) arc (180:0: 1em);
    \draw[black, thick, densely dashed] (3.5em, 0.em) arc (180:0: .5em);
    \draw[transparent, thick, densely dashed] (2em, 0.em) arc (180:360: 2em);
    \node at (2em, 0.em)[circle,fill,inner sep=.1em]{};
    \node at (3.5em, 0.em)[circle,fill,inner sep=.1em]{};
    \node at (3em, 0.em)[circle,fill,inner sep=.1em]{};
    \node at (4.5em, 0.em)[circle,fill,inner sep=.1em]{};
    \node at (5em, 0.em)[circle,fill,inner sep=.1em]{};
    \node at (6em, 0.em)[circle,fill,inner sep=.1em]{};
    \end{tikzpicture}
    &
    \begin{tikzpicture}
    \draw[black, thick] (1.em, 0.em) -- (7.em, 0.em);
    \draw[black, thick, densely dashed] (2em, 0.em) arc (180:0: 2em);
    \draw[black, thick, densely dashed] (3em, 0.em) arc (180:360: .75em);
    \draw[black, thick, densely dashed] (3.5em, 0.em) arc (180:0: .75em);
    \draw[transparent, thick, densely dashed] (2em, 0.em) arc (180:360: 2em);
    \node at (2em, 0.em)[circle,fill,inner sep=.1em]{};
    \node at (3.5em, 0.em)[circle,fill,inner sep=.1em]{};
    \node at (3em, 0.em)[circle,fill,inner sep=.1em]{};
    \node at (4.5em, 0.em)[circle,fill,inner sep=.1em]{};
    \node at (5em, 0.em)[circle,fill,inner sep=.1em]{};
    \node at (6em, 0.em)[circle,fill,inner sep=.1em]{};
    \end{tikzpicture}
     \\ (6) & (7) & (8) & (9) & (10) \\[1em]
     \begin{tikzpicture}
    \draw[black, thick] (1.em, 0.em) -- (7.em, 0.em);
    \draw[black, thick, densely dashed] (2em, 0.em) arc (180:0: 2em);
    \draw[black, thick, densely dashed] (4.25em, 0.em) arc (180:0: .625em);
    \draw[black, thick, densely dashed] (2.5em, 0.em) arc (180:0: .625em);
    \draw[transparent, thick, densely dashed] (2em, 0.em) arc (180:360: 2em);
    \node at (2em, 0.em)[circle,fill,inner sep=.1em]{};
    \node at (2.5em, 0.em)[circle,fill,inner sep=.1em]{};
    \node at (3.75em, 0.em)[circle,fill,inner sep=.1em]{};
    \node at (4.25em, 0.em)[circle,fill,inner sep=.1em]{};
    \node at (5.5em, 0.em)[circle,fill,inner sep=.1em]{};
    \node at (6em, 0.em)[circle,fill,inner sep=.1em]{};
    \end{tikzpicture}
    &
    \begin{tikzpicture}
    \draw[black, thick] (1.em, 0.em) -- (7.em, 0.em);
    \draw[black, thick, densely dashed] (2em, 0.em) arc (180:0: 1.5em);
    \draw[black, thick, densely dashed] (3em, 0.em) arc (180:360: 1.5em);
    \draw[black, thick, densely dashed] (3.5em, 0.em) arc (180:0: .5em);
    \draw[transparent, thick, densely dashed] (2em, 0.em) arc (180:360: 2em);
    \node at (2em, 0.em)[circle,fill,inner sep=.1em]{};
    \node at (3.5em, 0.em)[circle,fill,inner sep=.1em]{};
    \node at (3em, 0.em)[circle,fill,inner sep=.1em]{};
    \node at (4.5em, 0.em)[circle,fill,inner sep=.1em]{};
    \node at (5em, 0.em)[circle,fill,inner sep=.1em]{};
    \node at (6em, 0.em)[circle,fill,inner sep=.1em]{};
    \end{tikzpicture}
    &
    \begin{tikzpicture}
    \draw[black, thick] (1.em, 0.em) -- (7.em, 0.em);
    \draw[black, thick, densely dashed] (2em, 0.em) arc (180:0: 1.5em);
    \draw[black, thick, densely dashed] (2.5em, 0.em) arc (180:0: 1.em);
    \draw[black, thick, densely dashed] (3.5em, 0.em) arc (180:360: 1.25em);
    \draw[transparent, thick, densely dashed] (2em, 0.em) arc (180:360: 2em);
    \node at (2em, 0.em)[circle,fill,inner sep=.1em]{};
    \node at (3.5em, 0.em)[circle,fill,inner sep=.1em]{};
    \node at (2.5em, 0.em)[circle,fill,inner sep=.1em]{};
    \node at (4.5em, 0.em)[circle,fill,inner sep=.1em]{};
    \node at (5em, 0.em)[circle,fill,inner sep=.1em]{};
    \node at (6em, 0.em)[circle,fill,inner sep=.1em]{};
    \end{tikzpicture}
    &
    \begin{tikzpicture}
    \draw[black, thick] (1.em, 0.em) -- (7.em, 0.em);
    \draw[black, thick, densely dashed] (2em, 0.em) arc (180:0: 1.5em);
    \draw[black, thick, densely dashed] (2.5em, 0.em) arc (180:0: .5em);
    \draw[black, thick, densely dashed] (4.em, 0.em) arc (180:360: 1.em);
    \draw[transparent, thick, densely dashed] (2em, 0.em) arc (180:360: 2em);
    \node at (2em, 0.em)[circle,fill,inner sep=.1em]{};
    \node at (3.5em, 0.em)[circle,fill,inner sep=.1em]{};
    \node at (2.5em, 0.em)[circle,fill,inner sep=.1em]{};
    \node at (4.em, 0.em)[circle,fill,inner sep=.1em]{};
    \node at (5em, 0.em)[circle,fill,inner sep=.1em]{};
    \node at (6em, 0.em)[circle,fill,inner sep=.1em]{};
    \end{tikzpicture}
    &
    \begin{tikzpicture}
    \draw[black, thick] (1.em, 0.em) -- (7.em, 0.em);
    \draw[black, thick, densely dashed] (2em, 0.em) arc (180:0: 1.25em);
    \draw[black, thick, densely dashed] (3em, 0.em) arc (180:360: 1em);
    \draw[black, thick, densely dashed] (3.5em, 0.em) arc (180:0: 1.25em);
    \draw[transparent, thick, densely dashed] (2em, 0.em) arc (180:360: 2em);
    \node at (2em, 0.em)[circle,fill,inner sep=.1em]{};
    \node at (3.5em, 0.em)[circle,fill,inner sep=.1em]{};
    \node at (3em, 0.em)[circle,fill,inner sep=.1em]{};
    \node at (4.5em, 0.em)[circle,fill,inner sep=.1em]{};
    \node at (5em, 0.em)[circle,fill,inner sep=.1em]{};
    \node at (6em, 0.em)[circle,fill,inner sep=.1em]{};
    \end{tikzpicture}
    \\ (11) & (12) & (13) & (14) & (15) \\[1em]
    \begin{tikzpicture}
    \draw[black, thick] (1.em, 0.em) -- (7.em, 0.em);
    \draw[transparent, thick, densely dashed] (2em, 0.em) arc (180:0: 2em);
    \draw[black, thick, densely dashed] (2em, 0.em) arc (180:0: .75em);
    \draw[black, thick, densely dashed] (3em, 0.em) arc (180:360: 1em);
    \draw[black, thick, densely dashed] (4.5em, 0.em) arc (180:0: .75em);
    \draw[transparent, thick, densely dashed] (2em, 0.em) arc (180:360: 2em);
    \node at (2em, 0.em)[circle,fill,inner sep=.1em]{};
    \node at (3.5em, 0.em)[circle,fill,inner sep=.1em]{};
    \node at (3em, 0.em)[circle,fill,inner sep=.1em]{};
    \node at (4.5em, 0.em)[circle,fill,inner sep=.1em]{};
    \node at (5em, 0.em)[circle,fill,inner sep=.1em]{};
    \node at (6em, 0.em)[circle,fill,inner sep=.1em]{};
    \end{tikzpicture}
    \\ (16) 
  \end{tabular}
  \caption{Three-loop diagrams giving contributions to the fermion field anomalous dimensions, containing Yukawa and quartic interactions of fermions (solid lines) and scalar fields (dashed lines). }
  \label{fig:gamma_f}
  \end{figure}
  
  Similar to the scalar case, fermions are renormalised by introducing a field strength matrix $\psi_i \mapsto (\sqrt{Z^\psi})_{i}^{\ j} \psi_j$.
  Dropping the indices for convenience, the fermion anomalous dimension is defined via 
  \begin{equation}
    \gamma^\psi =  \frac{\mathrm{d} \sqrt{Z^\psi} }{\mathrm{d} \ln \mu} \left(\sqrt{Z^\psi}\right)^{-1} = \sum_{n=1}^\infty \frac{\gamma^{\psi, n\ell}}{(4\pi)^{2n}}
  \end{equation}
  Again, the leading and next-to-leading orders are available in full generality \cite{Machacek:1983tz,Luo:2002ti}. 
  Diagrams contributing to the three-loop expression $\gamma^{\psi,3\ell}$ containing Yukawa and quartic interactions are listed in \fig{gamma_f}. Here, (8), (13) and (14) are not symmetric under permutation of their external legs. 
  Similar to the scalar case $\sqrt{Z^\phi}$, these are linked to a potential ambiguity of an unitary transformation for fermionic field multiplets in the definition of $\sqrt{Z^\psi}$, which is unphysical, see \cite{Bednyakov:2014pia,Herren:2017uxn, Jack:2016tpp} for a more detailed discussion. In this work, we use \cite{Herren:2017uxn} to determine all parameters, electing anomalous dimensions to be hermitian. This fixes the ambiguous diagrams in \fig{gamma_f} to be symmetrised, thus leading to the general expression
  \begin{equation}\label{eq:gamma_f}
  \begin{aligned}
    \gamma^{\psi,3\ell} &= - \tfrac{11}{96}\,\lambda_{acde} \lambda_{bcde}\, (y^{ab}) + \lambda_{abcd}\, (y^{abcd})  \\
    &\phantom{=} - \tfrac3{32}\, \tr(y^{ac}) \, \tr(y^{bc}) \, (y^{ab}) + \tr( y^{abcc} + \tfrac12 y^{acbc} ) (y^{ab}) \\
    &\phantom{=} - \tfrac3{32} \tr(y^{ab}) \left[\tfrac43\, y^{accb}  - 3\, y^{cabc} + y^{acbc} + y^{cacb} \right] \\
   &\phantom{=} - \tfrac5{32} \, (y^{abbcca}) + \tfrac1{16} \, (y^{abccba})   - \tfrac5{16} \, (y^{abcbca})  + \tfrac14 \, (y^{abccab}) \\
   &\phantom{=}   + \tfrac3{32} \, (y^{abbcac}  + y^{abaccb}) + \left[\tfrac32 \zeta_3 - 1 \right] (y^{abcabc}) + \tfrac12\, (y^{abacbc}).
  \end{aligned}
\end{equation}
In particular, this means that the two diagrams depicted by \fig{gamma_f}~(13) do not contribute to the fermion anomalous dimension, while (8) and (14) are symmetrised.
\section{Yukawa Interaction and Fermion Masses}\label{sec:beta-yuk}
\begin{figure}
  \centering
\begin{tabular}{ccccccc}
    \begin{tikzpicture}
    \draw[black, thick] (2.em, 0.em) -- (4em, 2.em) -- (6.em, 0.em);
    \draw[black, thick, densely dashed] (4em, 4.em) -- (4em, 2.em);
    \draw[black, thick, densely dashed] (4em, .5em) circle (.6em);
    \draw[black, thick, densely dashed] (2.5em, .5em) -- (5.5em, .5em); 
    \node at (4em, 2.em)[circle,fill,inner sep=.1em]{};
    \node at (3.4em, .5em)[circle,fill,inner sep=.1em]{};
    \node at (4.6em, .5em)[circle,fill,inner sep=.1em]{};
    \node at (3.4em, .5em)[circle,fill,inner sep=.1em]{};
    \node at (2.5em, .5em)[circle,fill,inner sep=.1em]{};
    \node at (5.5em, .5em)[circle,fill,inner sep=.1em]{};
    \end{tikzpicture}
    &
    \begin{tikzpicture}
    \draw[black, thick] (2.em, 0.em) -- (2.5em, .5em) -- (5.5em, .5em) -- (6.em, 0.em);
    \draw[black, thick, densely dashed] (4em, 4.em) -- (4em, 1.85em) -- (2.5em, .5em);
    \draw[black, thick, densely dashed] (4.5em, 1.75em) circle (.5em);
    \draw[black, thick, densely dashed] (4em, .5em) -- (4.75em, 1.25em) -- (5.5em, .5em);
    \node at (4em, .5em)[circle,fill,inner sep=.1em]{};
    \node at (2.5em, .5em)[circle,fill,inner sep=.1em]{};
    \node at (5.5em, .5em)[circle,fill,inner sep=.1em]{};
    \node at (4.75em, 1.25em)[circle,fill,inner sep=.1em]{};
    \node at (4.em, 1.85em)[circle,fill,inner sep=.1em]{};
    \end{tikzpicture}
    &
    \begin{tikzpicture}
    \draw[black, thick] (2.em, 0.em) -- (2.5em, .5em) -- (5.5em, .5em) -- (6.em, 0.em);
    \draw[black, thick, densely dashed] (4em, 4.em) -- (4em, .5em);
    \draw[black, thick, densely dashed] (4.5em, 1.75em) circle (.5em);
    \draw[black, thick, densely dashed] (2.5em, .5em) -- (4.75em, 1.25em) -- (5.5em, .5em);
    \node at (4em, .5em)[circle,fill,inner sep=.1em]{};
    \node at (2.5em, .5em)[circle,fill,inner sep=.1em]{};
    \node at (5.5em, .5em)[circle,fill,inner sep=.1em]{};
    \node at (4.75em, 1.25em)[circle,fill,inner sep=.1em]{};
    \node at (4.em, 1.85em)[circle,fill,inner sep=.1em]{};
    \end{tikzpicture}
    &
    \begin{tikzpicture}
    \draw[black, thick] (2.em, 0.em) -- (2.5em, .5em) -- (5.5em, .5em) -- (6.em, 0.em);
    \draw[black, thick, densely dashed] (4em, 4.em) -- (4em, 2.5em) -- (2.5em, .5em);
    \draw[black, thick, densely dashed] (4em, 1.65em) -- (4em, 2.5em) -- (5.5em, .5em);
    \draw[black, thick, densely dashed] (4em, .95em) -- (4em, .5em);
    \draw[black, thick] (4em, 1.3em) circle (.35em);
    \node at (4em, .5em)[circle,fill,inner sep=.1em]{};
    \node at (2.5em, .5em)[circle,fill,inner sep=.1em]{};
    \node at (5.5em, .5em)[circle,fill,inner sep=.1em]{};
    \node at (4.em, 2.5em)[circle,fill,inner sep=.1em]{};
    \node at (4.em, 1em)[circle,fill,inner sep=.1em]{};
    \node at (4.em, 1.6em)[circle,fill,inner sep=.1em]{};
    \end{tikzpicture}
    &
    \begin{tikzpicture}
    \draw[black, thick] (2.em, 0.em) -- (2.5em, .5em) -- (5.5em, .5em) -- (6.em, 0.em);
    \draw[black, thick, densely dashed] (4em, 4.em) -- (4em, 2.5em) -- (2.5em, .5em);
    \draw[black, thick, densely dashed] (4em, .5em) -- (4em, 2.5em) -- (5.5em, .5em);
    \draw[black, thick, fill=white] (4.9em, 1.3em) circle (.35em);
    \node at (4em, .5em)[circle,fill,inner sep=.1em]{};
    \node at (2.5em, .5em)[circle,fill,inner sep=.1em]{};
    \node at (5.5em, .5em)[circle,fill,inner sep=.1em]{};
    \node at (4.em, 2.5em)[circle,fill,inner sep=.1em]{};
    \node at (4.7em, 1.6em)[circle,fill,inner sep=.1em]{};
    \node at (5.1em, 1.em)[circle,fill,inner sep=.1em]{};
    \end{tikzpicture}
    &
    \begin{tikzpicture}
    \draw[black, thick] (2.em, 0.em) -- (2.5em, .5em) -- (4em, 2em) -- (5.5em, .5em) -- (6.em, 0.em);
    \draw[black, thick, densely dashed] (4em, 4.em) -- (4em, 2.em);
    \draw[black, thick, densely dashed] (2.5em, .5em)  -- (5.5em, .5em);
    \draw[black, thick, densely dashed] (3.25em, 1.25em)  -- (4em, .5em) -- (4.75em, 1.25em);
    \node at (4em, 2em)[circle,fill,inner sep=.1em]{};
    \node at (2.5em, .5em)[circle,fill,inner sep=.1em]{};
    \node at (5.5em, .5em)[circle,fill,inner sep=.1em]{};
    \node at (3.25em, 1.25em)[circle,fill,inner sep=.1em]{};
    \node at (4.75em, 1.25em)[circle,fill,inner sep=.1em]{};
    \node at (4em, .5em)[circle,fill,inner sep=.1em]{};
    \end{tikzpicture}
    &
    \begin{tikzpicture}
    \draw[black, thick] (2.em, 0.em) -- (2.5em, .5em) -- (4em, 2em) -- (5.5em, .5em) -- (6.em, 0.em);
    \draw[black, thick, densely dashed] (4em, 4.em) -- (4em, 2.em);
    \draw[black, thick, densely dashed] (2.5em, .5em)  -- (5.5em, .5em);
    \draw[black, thick, densely dashed] (5em, 1.em)  -- (4em, .5em) -- (4.5em, 1.5em);
    \node at (4em, 2em)[circle,fill,inner sep=.1em]{};
    \node at (2.5em, .5em)[circle,fill,inner sep=.1em]{};
    \node at (5.5em, .5em)[circle,fill,inner sep=.1em]{};
    \node at (4.5em, 1.5em)[circle,fill,inner sep=.1em]{};
    \node at (5em, 1.em)[circle,fill,inner sep=.1em]{};
    \node at (4em, .5em)[circle,fill,inner sep=.1em]{};
    \end{tikzpicture}
    \\
\end{tabular}
  \caption{Yukawa proper vertex diagrams at three-loop order, including scalar (dashed lines) and fermionic propagators (solid lines).}
  \label{fig:yuk}
\end{figure}
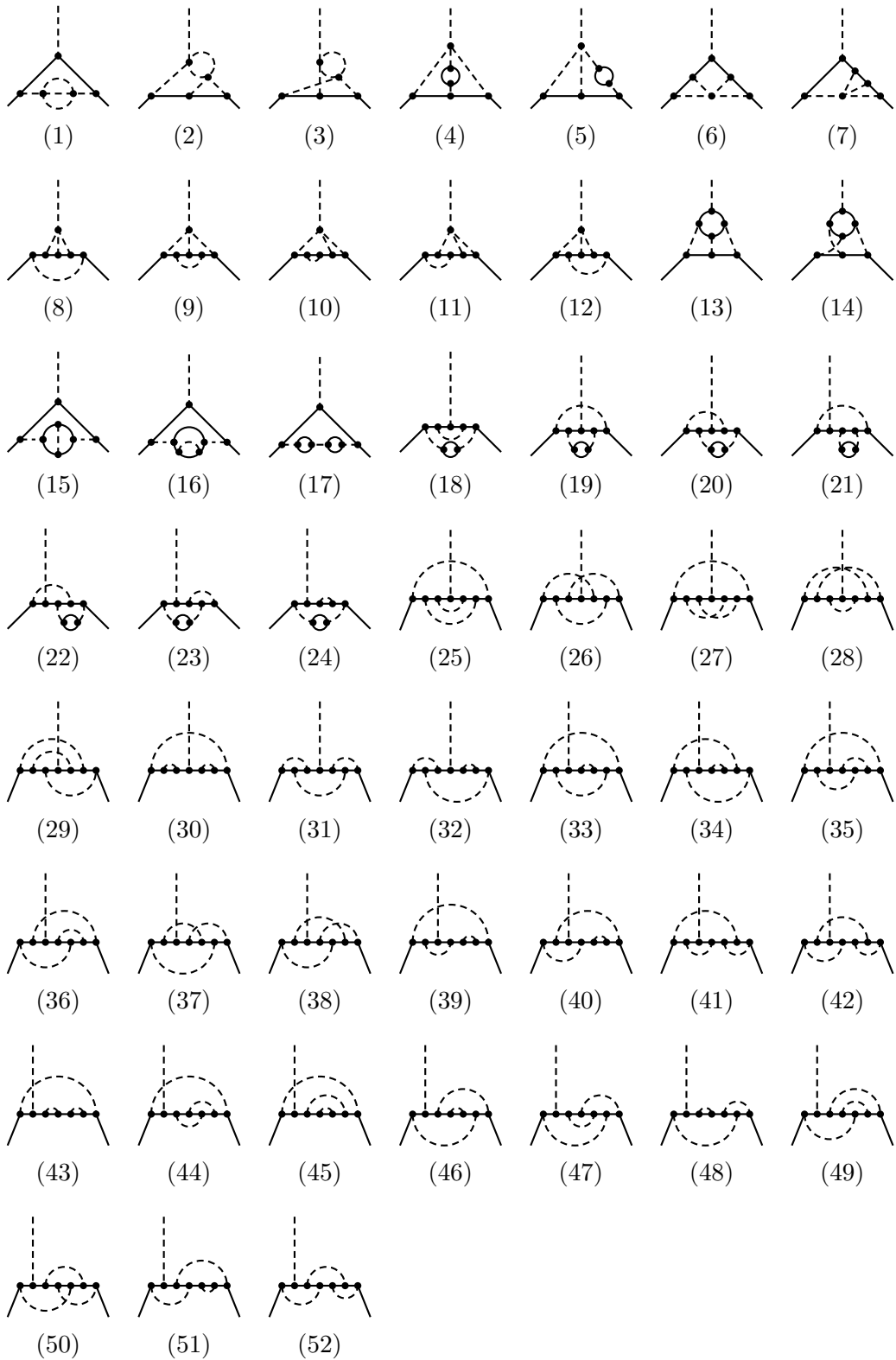
The $\beta$-functions of the renormalised Yukawa couplings $y^{aij}$ are defined via 
\begin{equation}
  \beta_{y^a} = \frac{\mathrm{d} \,y^a}{\mathrm{d} \ln \mu} = \sum_{n=1}^\infty (4\pi)^{-2n} \beta_{y^a}^{n\ell},
\end{equation} 
and template formulas for the one- and two-loop expressions can be found in \cite{Machacek:1983fi,Luo:2002ti,Schienbein:2018fsw,Poole:2019kcm}.
The three-loop part $\beta_{y^a}^{3\ell}$ consists of fermionic \eq{gamma_f} and scalar leg contributions \eq{gamma_s}, as well as proper Yukawa vertex corrections. In our setup without gauge interactions, the latter are listed in \fig{yuk} and give rise to 52 open parameters, considering that the vertex is symmetric under permutation of its fermionic legs. This ansatz has been cross-checked with diagram list provided with \cite{Poole:2019kcm}. Using \cite{Herren:2017uxn}, the three-loop expression
\begin{equation}\label{eq:beta_yuk}
\begin{aligned}
  \beta_{y^a}^{3\ell} &= \gamma_{ab}^{\phi,3\ell} y^b +  \gamma^{\psi,3\ell} y^a + y^a \left(\gamma^{\psi,3\ell}\right)^\intercal - \tfrac38 \lambda_{bdef} \lambda_{cdef}\,(y^{bac}) \\
  &\phantom{=}  + \tfrac12\left[ \lambda_{aebf} \lambda_{bfcd}  +  \lambda_{abcd} \, \tr(y^{be})\right] \left[y^{cde} + y^{ecd}  +  3\,y^{ced} \right] \\
  &\phantom{=} + \lambda_{bcde}\,(2\,y^{bcade} + 3\, y^{bacde} + 3\, y^{bcdae}) \\
  &\phantom{=} + \lambda_{abcd} \left[ 5 \, y^{ebcde} + 3 \, y^{beced}   + \tfrac12\, y^{beecd}  + \tfrac12\, y^{bceed} \right] \\
  &\phantom{=} + \lambda_{abcd} \left[ - y^{ebecd} - y^{bcede} + 2\, y^{ebced} + 2\, y^{becde} \right] - \tfrac12\, \tr(y^{bd}) \, \tr(y^{dc}) (y^{bac})\\
  &\phantom{=} + 2 \left[3 \zeta_3 - 2\right] \, \tr(y^{abcd}) \left[ y^{bdc} + y^{cbd} \right] +  \left[\tfrac54\tr(y^{bdcd}) + \tfrac{25}8 \,\tr(y^{bcdd}) \right] (y^{bac})  \\
  &\phantom{=} + \tr(y^{bc}) \left[ 2\, y^{dbacd} - y^{bdadc} - \tfrac12\, y^{bdacd} - \tfrac12\, y^{dbadc} + \tfrac{25}{16} \, y^{dbcad} + \tfrac{25}{16} \,y^{dabcd}\right] \\
    &\phantom{=} + \tfrac32\,\tr(y^{bc}) \left[   y^{bdcad} + y^{dabdc} - y^{dbdac} -  y^{badcd} - \tfrac13  y^{bddac} - \tfrac13  y^{baddc}\right] \\
     &\phantom{=} + 4\, y^{bcdadcb} - 3 \, y^{bcdadbc} + \left[6 \zeta_3 - 5\right]\,y^{bcdacdb} +\left[6 \zeta_3 - 2\right]\left[ y^{bcdacbd} + y^{bcdabdc} \right] \\
     &\phantom{=} + 2\, y^{bcdabcd}  - 2\, y^{bcbadcd} - \tfrac12\, y^{bccaddb} - \tfrac32\left[ y^{bcbaddc} + y^{bddacbc}\right] \\
     &\phantom{=} + y^{bc} (y^{add} + y^{dda}) (2\,y^{cb} - \tfrac32\,y^{bc}) - \,y^{bc} (y^{adc} + y^{dca} ) y^{db}\\
     &\phantom{=} + \left[6 \zeta_3- 3\right] ( y^{bcadbdc} + y^{bdcdabc}) - 4\, (y^{bcadcbd} + y^{dbcdacb}) - y^{bcacddb}  - y^{bddcacb}\\
      &\phantom{=} + \left[6 \zeta_3 - 2\right] ( y^{bcadbcd} + y^{dbcdabc}) - \tfrac12\, ( y^{bcabddc} +  y^{bddcabc} )  - 3\,(y^{bcacdbd} + y^{bdbcacd}) \\
      &\phantom{=} + y^{bc} (y^{abd} + y^{bda} ) y^{cd} - \tfrac12(y^{baccddb} + y^{bccddab}) - y^{bacdcdb} - y^{bcdcdab} \\
      &\phantom{=} + \tfrac7{16} (y^{bacddcb} + y^{bcddcab})  + \tfrac12 ( y^{bacddbc}  + y^{bcddbac} )  - 2( y^{badcdbc} + y^{bcdbdac} ) \\
      &\phantom{=} - \tfrac32  (y^{baddcbc} + y^{bcbddac}) + y^{bacdbdc} + y^{bdcdbac} + 3\left[2 \zeta_3 - 1\right] ( y^{badcbdc} + y^{bdcbdac}) \\
      & \phantom{=} + \tfrac32  (y^{bacbddc}  + y^{bddcbac}) + 2(y^{badbcdc} + y^{bdbcdac})
\end{aligned}
\end{equation}
is obtained. Note that the contribution of \fig{yuk}(13) in \eq{beta_yuk} vanishes. This automatically provides the renormalisation group evolution for the fermion mass term $\beta_{\mathrm{m}}^{3\ell}$ after the replacement
\begin{equation}
  \gamma_{ab}^{\phi,3\ell} y^b \mapsto 0, \qquad \qquad  y^a \mapsto \mf, \qquad \qquad \lambda_{abcd} \mapsto h_{bcd}
\end{equation}
in \eq{beta_yuk}. Furthermore, the result \eq{beta_yuk} has been cross-checked against \cite{Bednyakov:2012en,Chetyrkin:2012rz,Jack:2013sha,Bednyakov:2014pia}. For convenience, we have collected all coefficients of $\beta_{y}^{3\ell}$ in the basis from \cite{Poole:2019kcm} in Appendix~\ref{app:WeylCC}.

\section{Scalar Quartic Interactions}\label{sec:beta-quart}

\begin{figure}[ht]
  \centering
\begin{tabular}{ccccc}
    \begin{tikzpicture}
    \draw[black, thick, densely dashed]  (1.em, 1em) -- (2em, 0em) -- (1em, -1em);
    \draw[black, thick, densely dashed]  (4em,0em) circle (2em);
    \draw[black, thick, densely dashed]  (5.6em, 1.2em) to [bend right=60] (5.6em, -1.2em);
    \draw[black, thick, densely dashed]  (5.6em, 1.2em) -- (6.6em, 1.7em);
    \draw[black, thick, densely dashed]  (5.6em, -1.2em) -- (6.6em, -1.7em);
    \draw[black, thick, fill=white] (5em, 0em) circle (.5em);
    \draw[transparent, thick] (4em, -2em) circle (.5em);
    \node at (2em, 0em)[circle,fill,inner sep=.1em]{};
    \node at (5.6em, 1.2em)[circle,fill,inner sep=.1em]{};
    \node at (5.6em, -1.2em)[circle,fill,inner sep=.1em]{};
    \node at (5.07em, .47em)[circle,fill,inner sep=.1em]{};
    \node at (5.07em, -.47em)[circle,fill,inner sep=.1em]{};
    \end{tikzpicture} 
    &
    \begin{tikzpicture}
    \draw[black, thick, densely dashed]  (1.em, 1em) -- (2em, 0em) -- (1em, -1em);
    \draw[black, thick, densely dashed]  (4em,0em) circle (2em);
    \draw[black, thick, densely dashed]  (5.6em, 1.2em) to [bend right=60] (5.6em, -1.2em);
    \draw[black, thick, densely dashed]  (5.6em, 1.2em) -- (6.6em, 1.7em);
    \draw[black, thick, densely dashed]  (5.6em, -1.2em) -- (6.6em, -1.7em);
    \draw[black, thick, fill=white] (4em, 2em) circle (.5em);
    \draw[transparent, thick] (4em, -2em) circle (.5em);
    \node at (2em, 0em)[circle,fill,inner sep=.1em]{};
    \node at (5.6em, 1.2em)[circle,fill,inner sep=.1em]{};
    \node at (5.6em, -1.2em)[circle,fill,inner sep=.1em]{};
    \node at (3.5em, 1.9em)[circle,fill,inner sep=.1em]{};
    \node at (4.5em, 1.9em)[circle,fill,inner sep=.1em]{};
    \end{tikzpicture}
    &
    \begin{tikzpicture}
    \draw[black, thick, densely dashed]  (1.em, 1em) -- (2em, 0em) -- (1em, -1em);
    \draw[black, thick, densely dashed]  (7.em, 1em) -- (6em, 0em) -- (7em, -1em);
    \draw[black, thick, densely dashed]  (4em,0em) circle (2em);
    \draw[black, thick, fill=white] (4em, 2em) circle (.5em);
    \draw[black, thick, fill=white] (4em, -2em) circle (.5em);
    \node at (2em, 0em)[circle,fill,inner sep=.1em]{};
    \node at (6em, 0em)[circle,fill,inner sep=.1em]{};
    \node at (3.5em, 1.9em)[circle,fill,inner sep=.1em]{};
    \node at (4.5em, 1.9em)[circle,fill,inner sep=.1em]{};
    \node at (3.5em, -1.9em)[circle,fill,inner sep=.1em]{};
    \node at (4.5em, -1.9em)[circle,fill,inner sep=.1em]{};
    \end{tikzpicture}
    &
    \begin{tikzpicture}
    \draw[black, thick, densely dashed]  (1.em, 1em) -- (2em, 0em) -- (1em, -1em);
    \draw[black, thick, densely dashed]  (7.em, 1em) -- (6em, 0em) -- (7em, -1em);
    \draw[black, thick, densely dashed]  (4em,0em) circle (2em);
    \draw[black, thick, fill=white] (5.2em, 1.4em) circle (.5em);
    \draw[black, thick, fill=white] (2.8em, 1.4em) circle (.5em);
    \draw[transparent, thick] (4em, -2em) circle (.5em);
    \node at (2em, 0em)[circle,fill,inner sep=.1em]{};
    \node at (6em, 0em)[circle,fill,inner sep=.1em]{};
    \node at (3.1em, 1.8em)[circle,fill,inner sep=.1em]{};
    \node at (4.9em, 1.8em)[circle,fill,inner sep=.1em]{};
    \node at (2.4em, 1.1em)[circle,fill,inner sep=.1em]{};
    \node at (5.6em, 1.1em)[circle,fill,inner sep=.1em]{};
    \end{tikzpicture}
    &
    \begin{tikzpicture}
    \draw[black, thick, densely dashed]  (1.em, 1em) -- (2em, 0em) -- (1em, -1em);
    \draw[black, thick, densely dashed]  (7.em, 1em) -- (6em, 0em) -- (7em, -1em);
    \draw[black, thick, densely dashed]  (4em,0em) circle (2em);
    \draw[black, thick, fill=white] (4em, -1.5em) circle (1em);
    \draw[black, thick, densely dashed] (4em, -0.5em) -- (4em, -2.5em);
    \node at (2em, 0em)[circle,fill,inner sep=.1em]{};
    \node at (6em, 0em)[circle,fill,inner sep=.1em]{};
    \node at (6em, 0em)[circle,fill,inner sep=.1em]{};
    \node at (3.05em, -1.72em)[circle,fill,inner sep=.1em]{};
    \node at (4.95em, -1.72em)[circle,fill,inner sep=.1em]{};
    \node at (4em, -0.5em)[circle,fill,inner sep=.1em]{};
    \node at (4em, -2.5em)[circle,fill,inner sep=.1em]{};
    \end{tikzpicture}
    \\
\end{tabular}
  \caption{Three-loop diagrams contributing to the scalar quartic vertex renormalisation, containing both quartic and Yukawa vertices. Solid [dashed] lines correspond to fermionic [scalar] propagators. The number of inequivalent permutations of the external legs for each diagram is also indicated. }
  \label{fig:quart1}
\end{figure}
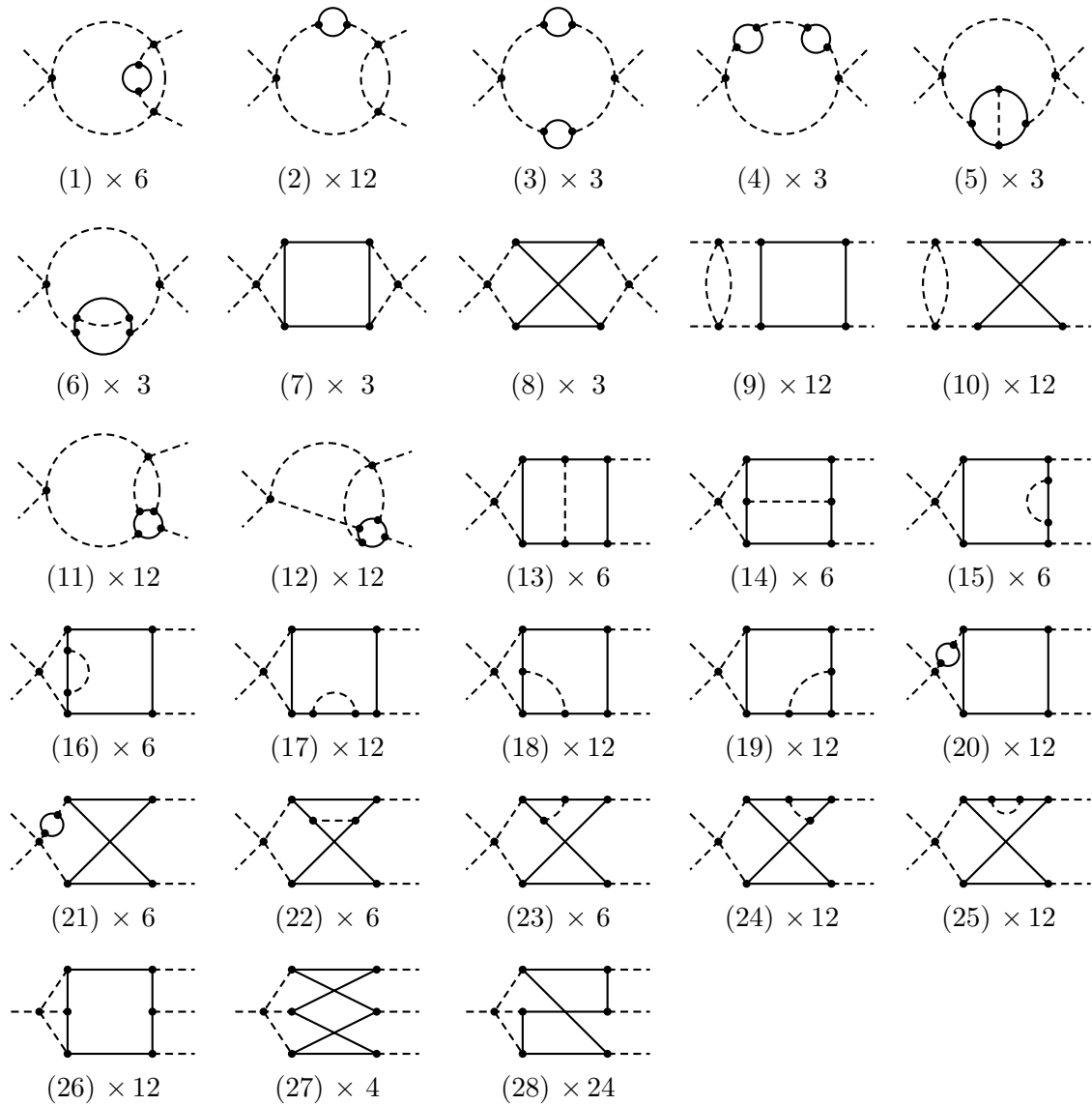
\begin{figure}[ht]
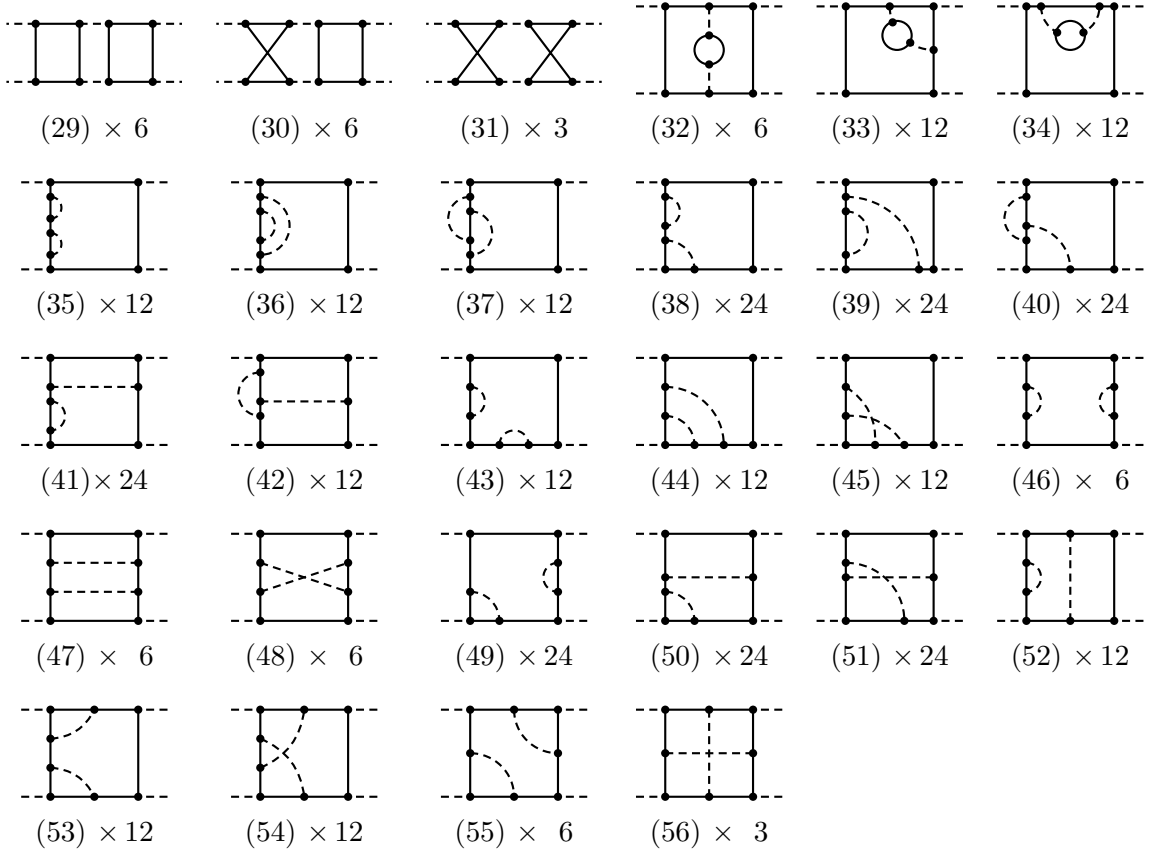

  \centering
  %
  \caption{Three-loop diagrams contributing to the scalar quartic vertex renormalisation, containing only Yukawa couplings with fermion (solid lines) and scalar propagators (dashed lines) propagators. The number of inequivalent permutations of external legs is listed below each diagram.}
  \label{fig:quart2}
\end{figure}
For scalar quartic interactions $\lambda_{abcd}$, the $\beta$-functions
\begin{equation}
  \beta_{\lambda_{abcd}} = \frac{\mathrm{d} \lambda_{abcd}}{\mathrm{d} \ln \mu} = \sum_{n=1}^\infty (4\pi)^{-2n} \beta_{\lambda_{abcd}}^{n\ell}  
\end{equation} 
have been computed for general QFTs at two-loop order \cite{Machacek:1984zw,Luo:2002ti,Schienbein:2018fsw,Poole:2019kcm}, and will now be extended to three-loop in the gaugeless limit via the ansatz \eq{master-template}.
These expressions contain scalar leg corrections \eq{gamma_s} which have been already determined. Moreover, the expression features proper scalar quartic vertex corrections that reflect the overall symmetrisation of the quartic interaction terms. These are listed in \fig{quart1} and \fig{quart2}, not including the six pure quartic contributions given in \cite{Jack:1990eb,Jack:2018oec,Steudtner:2020tzo}. Thus, 56 open parameters are implied. 
Three-loop SM results \cite{Chetyrkin:2012rz,Bednyakov:2013eba,Chetyrkin:2013wya,Bednyakov:2013cpa,Bednyakov:2014pia} as well as  $\beta$- and $\gamma$-functions in Gross-Neveu-Yukawa theories, in particular chiral Ising, XY and Heisenberg models \cite{Zerf:2017zqi,Mihaila:2017ble}, can be utilised for matching unknown coefficients. 
Additional input data can be extracted when comparing to $\mathcal{N} = 1$ supersymmetric QFTs as detailed in Appendix~\ref{app:susy}. 
Overall, from the 56 open parameters, 29 coefficients can be determined directly, and 14 supplementary conditions apply, leaving only \openparams{} pieces of information missing. 

We are currently unaware of pre-existing calculation that would fix these open coefficients. No data from the THDM is available, as \cite{Herren:2017uxn} does not include three-loop quartic $\beta$ functions, while \cite{Bednyakov:2018cmx} lacks the Yukawa contributions in question. Moreover, no conditions in \cite{Poole:2019kcm} apply to coefficients at this loop order. Similarily, no three-loop quartic RGEs are included in \cite{Jack:2013sha} either.

 Concretely, we obtain the three-loop $\beta$-function of the scalar quartic as
\begin{equation}\label{eq:quartic-sum}
  \begin{aligned}
    \beta_{\lambda_{abcd}}^{3\ell} &= \gamma_{ae}^{\phi,3\ell} \lambda_{ebcd} + \gamma_{be}^{\phi,3\ell} \lambda_{aecd} + \gamma_{ce}^{\phi,3\ell} \lambda_{abed} + \gamma_{de}^{\phi,3\ell} \lambda_{abce} \\
     &\phantom{=\,} + \beta_{abcd}^{\lambda:\,\lambda^4} + \beta_{abcd}^{\lambda:\,\lambda^3 y^2} + \beta_{abcd}^{\lambda:\,\lambda^2 y^4} + \beta_{abcd}^{\lambda:\,\lambda y^6} + \beta_{abcd}^{\lambda:\, y^8} 
  \end{aligned}
\end{equation}
where the pure scalar part has been computed earlier \cite{Jack:1990eb,Jack:2018oec,Steudtner:2020tzo} and reads
\begin{equation}
  \begin{aligned}
    \beta_{abcd}^{\lambda:\,\lambda^4} &= 12 \zeta_3 \,\lambda_{aefg} \lambda_{behi} \lambda_{cfhj} \lambda_{dgij} 
     - \tfrac12 \left[ \lambda_{aefg} \lambda_{bfgh} \lambda_{ceij} \lambda_{dhij} +  \perm{5}\right] \\
     &\phantom{=\,} + \tfrac12 \left[ \lambda_{abef} \lambda_{egij} \lambda_{fhij} \lambda_{cdgh} +  \perm{2}\right] 
    - \tfrac38 \left[ \lambda_{abef} \lambda_{ehij} \lambda_{ghij} \lambda_{cdfg} +  \perm{2}\right]\\
    &\phantom{=\,} - \tfrac12 \left[ \lambda_{abef} \lambda_{cegh} \lambda_{ghij} \lambda_{dfij} +  \perm{5}\right] + 2 \left[ \lambda_{abef}  \lambda_{cegh} \lambda_{fgij} \lambda_{dhij} + \perm{11}\right].\\
  \end{aligned}
\end{equation}
Insertions of a single Yukawa bubble $\tr(y^2)$ give rise to the correction
\begin{equation}
  \begin{aligned}
    \beta_{abcd}^{\lambda:\,\lambda^3 y^2} &= 2 \left[ \lambda_{abef} \lambda_{cegh} \lambda_{dfgi} \, \tr(y^{hi}) + \perm{5} \right] \qquad \qquad \qquad \qquad \qquad \qquad \qquad \\ 
    &\phantom{=\,}  -  \tfrac12  \left[ \lambda_{abef} \lambda_{cegh} \lambda_{dghi} \, \tr(y^{fi}) + \perm{11} \right].
  \end{aligned}
\end{equation}
Further, $\lambda^2\,\tr(y^2)\,\tr(y^2)$ and $\lambda^2\,\tr(y^4)$ terms contribute to the $\beta$-function via
\begin{equation}
  \begin{aligned}
    \beta_{abcd}^{\lambda:\,\lambda^2 y^4} &= - \tfrac14 \left[\lambda_{abef} \lambda_{cdgh}  + \perm{2}\right] \left[\,\tr(y^{eg}) \,\tr(y^{fh}) + 2\,\delta^{eg} \,\tr(y^{fi})\,\tr(y^{ih})\right]\\
    &\phantom{=\,} + \left[ \lambda_{abef} \lambda_{cdgh} + \perm{2} \right] \left[(3\zeta_3 - 1) \, \tr(y^{egfh}) + 2\, \tr(y^{efgh})\right] \\
    &\phantom{=\,} + \left[ \lambda_{abef} \lambda_{cdeg} + \perm{2} \right] \left[\tfrac{25}8\, \tr(y^{fghh}) + \tfrac54 \, \tr(y^{fhgh}) \right]\\
    &\phantom{=\,} +\left[ \lambda_{aefg} \lambda_{befh} \left(\tr(y^{cgdh}) + 3\,\tr(y^{cdgh}) \right)+ \perm{11}\right] \\
    &\phantom{=\,}+ 2 \left[ \lambda_{abef} \lambda_{cegh} \left( \tr(y^{dfgh})  + (3 \zeta_3 -2 ) \,\tr(y^{dgfh}) \right)+ \perm{11}\right].
  \end{aligned}
\end{equation}
While terms of the form $\lambda\,\tr(y^2)^3$ are absent, $\lambda\,\tr(y^2) \,\tr(y^4)$ and $\lambda\,\tr(y^6)$ contractions read
\begin{equation}\label{eq:l-ly6}
  \begin{aligned}
    \beta_{abcd}^{\lambda:\,\lambda y^6} &= \left[ \lambda_{abef}  \left(c_{(13)} \, \tr(y^{cdgefg})  -  10 \, \tr(y^{cgdegf})\right) + \perm{5}\right] \\
    &\phantom{=\,} -  \left[ \lambda_{abef} \left(  (5 + c_{(13)}) \,\tr(y^{cggdef}) + \tr(y^{cdeggf}) \right)+ \perm{5}\right] \\
    &\phantom{=\,} +\left[ \lambda_{abef} \,\tr\left(  - 3\, y^{cdggef} + c_{(18)} \, y^{cdgegf} \right)+ \perm{11}\right] \\
    &\phantom{=\,} + \left[ \lambda_{abef}\,\left(   - (8 + c_{(18)}) \,\tr(y^{cgdgef}) + 3\,\tr(y^{fg}) \,\tr(y^{gecd})\right) + \perm{11}\right] \\
    &\phantom{=\,} + 2\left[\lambda_{abef} \left( \tr(y^{fg}) \,\tr(y^{gced})  - ( 5 - 6\zeta_3 ) \tr(y^{cegdfg})\right)+ \perm{5} \right] \\
    &\phantom{=\,}   - [16 - 12 \zeta_3 + 2\,c_{(24)} ] \left[ \lambda_{abef}  \,\tr(y^{cgegdf}) + \perm{5}\right] \\
    &\phantom{=\,} + \left[ \lambda_{abef}\,\left( c_{(24)} \,\tr(y^{cegdgf})  -4\, \tr(y^{ecfdgg}) \right) + \perm{11}\right] \\
    &\phantom{=\,} + ( c_{(13)} - 8 )\left[ \lambda_{aefg}  \,\tr(y^{bcdefg})  + \perm{11} \right] \\
    &\phantom{=\,} - 24 \zeta_3 \left[ \lambda_{aefg}  \,\tr(y^{becfdg})  + \perm{3} \right] \\
    &\phantom{=\,} - \left(1 + 12 \zeta_3 + \tfrac12 c_{(13)} \right) \left[ \lambda_{aefg}  \,\tr(y^{bcedfg})  + \perm{23} \right],
  \end{aligned}
\end{equation}
where  the coefficients corresponding to contractions \fig{quart1}~(13), (18) and (24) could not be determined completely, and are retained as $c_{(13)}$, $c_{(18)}$  and $c_{(24)}$.
Finally, pure Yukawa terms $\tr(y^4)\,\tr(y^4)$, $\tr(y^2)\,\tr(y^6)$ and $\tr(y^8)$ are given by
\begin{equation}\label{eq:l-y8}
  \begin{aligned}
    \beta_{abcd}^{\lambda:\, y^8} &= - 8\left[\tr(y^{abef} + \tfrac32\,y^{aebf} ) \,\tr(y^{cdef}) + \perm{5} \right] \\
        &\phantom{=\,} - 4\left[\tr(y^{aebf}) \,\tr(y^{cedf}) + \perm{2} \right]  - 3  \left[\tr(y^{abecdf})\, \tr(y^{ef}) + \perm{5} \right] \\
     &\phantom{=\,} - \tr(y^{ef}) \left[4 \, \tr(y^{abcedf})+ \tfrac{25}8 \, \tr(y^{abcdef})\,  + \perm{11} \right] \\
    &\phantom{=\,} + \left[\tr(y^{abcdeeff}) +  \perm{11} \right] + \left[\tr(y^{abeecdff}) + \perm{5} \right] \\
    &\phantom{=\,} + \left[ - \tfrac78\, \tr(y^{abcdeffe}) + c_{(37)}\,\tr(y^{abcdefef}) + \perm{11}\right] \\
    &\phantom{=\,} +\left[ c_{(38)}\, \tr(y^{affebecd}) + c_{(39)}\, \tr(y^{aeffbecd}) +\perm{23}\right]\\
    &\phantom{=\,} +\left[ c_{(40)}\, \tr(y^{aefebfcd}) -  \tr(y^{affebced})  +\perm{23}\right]\\
    &\phantom{=\,} +\left[ c_{(42)}\, \tr(y^{aefebcfd}) +  \tr(y^{aeebffcd}) +\perm{11}\right]\\
    &\phantom{=\,} +\left[ c_{(44)}\, \tr(y^{aefbfecd}) + ( 8 - 12 \zeta_3 - c_{(13)} )\, \tr(y^{aefbefcd}) +\perm{11}\right]\\
    &\phantom{=\,} +\left[ -(8 + 2\,c_{(13)})\, \tr(y^{aefbcfed}) + c_{(48)}\, \tr(y^{aefbcefd}) +\perm{5}\right]\\
    &\phantom{=\,} +\left[ c_{(49)}\,\tr(y^{aebecffd})+ c_{(50)}\, \tr(y^{aefbfced}) +c_{(51)}\, \tr(y^{aefbecfd}) +\perm{23}\right]\\
    &\phantom{=\,} +(1 + c_{(13)}) \left[  \tr(y^{abecffde} ) + \perm{11}\right]\\
    &\phantom{=\,} +\left[ c_{(53)}\, \tr(y^{abecefdf} ) +(2+  c_{(13)})\, \tr(y^{abecfedf})  + \perm{11}\right]\\
    &\phantom{=\,} +c_{(55)}\,\left[ \tr(y^{aebfcedf}) + \perm{2}\right]  + c_{(56)}\,\left[  \tr(y^{aebecfdf})  +\perm{5}\right].\\
  \end{aligned}
\end{equation}
while no one-particle-irreducible diagrams are formed by $\tr(y^2)^4$ bubbles. For notational compactness, we have not inserted the relations
\begin{equation}
  \begin{aligned}
    c_{(51)} &=  - 6 (1+2\zeta_3) + 2 c_{(37)} + 3 c_{(38)} + c_{(39)} - c_{(40)} + c_{(42)} - c_{(44)} + 3 \,c_{(49)} +  3\,c_{(50)} \,,\\
    c_{(53)} &= 8 + 2 c_{(49)} + 2 c_{(50)} - c_{(55)}  \,,\\
    c_{(55)} &= -12 + 8 c_{(37)} + 12 c_{(38)} + 4 c_{(39)} + 4 c_{(42)} - 2 c_{(44)}  + 16 c_{(49)} + 16 c_{(50)}  \,,\\
    c_{(56)} &=  -16 (2 +3\zeta_3) - 4 c_{(37)} - 8 c_{(38)} - 8 c_{(39)} - 4 c_{(42)} - 2 c_{(48)} - 8 c_{(49)} - c_{(50)}  \,.
  \end{aligned}
\end{equation}
It may be possible to recover additional restrictions when comparing diagrams (38) with (49), (44) with (53) and (55), as well as (41) with (52) in \fig{quart2} with a more sophisticated analysis of momentum integrals.
Similar relations arise between diagrams (35), (43) and (46), as well as (3) and (4) in \fig{quart1}. However, this is insufficient to resolve all open parameters and will not be attempted here.

Just as in the Yukawa case, the renormalisation group equations for scalar cubic couplings $h_{abc}$ and scalar masses $m^2_{ab}$ can be obtained from eq.~\eq{quartic-sum}ff. using the dummy field method \cite{Martin:1993zk,Luo:2002ti,Schienbein:2018fsw}. This entails substituting external scalar indices by dummy ones: $\varnothing$, corresponding to a classical, non-propagating and uncharged singlet. Using the substitutions
\begin{equation} \label{eq:dummy}
  \lambda_{abc\varnothing} = h_{abc}, \qquad \lambda_{ab\varnothing\varnothing} = 2\,m^2_{ab}, \qquad y^{\varnothing} = \mf, \qquad   \gamma_{\varnothing e}^{\phi,3\ell} = 0
\end{equation}
on both sides of eq.~\eq{quartic-sum}ff. allows us to extract RGE parameters with positive canonical mass dimension. The last condition in \eq{dummy} reflects that there is no field strength renormalisation for the dummy field. 
 In general, the procedure will generate several distinct terms from permutations of the external indices in $\beta_{\lambda_{abcd}}$. For instance, the correction
\begin{equation}
  \beta_{\lambda_{abcd}} = ... + c_{(23)} \left[ \lambda_{abef}  \,\tr(y^{cegdgf})  + \perm{11} \right] + ...
\end{equation} from \eq{l-ly6} will lead to a contribution to the scalar mass RG evolution that reads
\begin{equation}
  \begin{aligned}
    \beta_{m^2_{ab}} &= ... + c_{(23)} \lambda_{abef} \,\tr(\mf \, y^{eg} \mf\, y^{gf}) + c_{(23)} \, m^2_{ef} \, \,\tr(y^{aegbgf} + y^{begagf}) \\
      &\phantom{=\, } + c_{(23)} \, h_{aef} \, \tr(\mf \,y^{egbgf} + y^{beg}\mf\,y^{gf}) + c_{(23)}\, h_{bef}\,  \tr(\mf\, y^{egagf} + y^{aeg}\mf\,y^{gf}) 
  \end{aligned}
\end{equation}
which is significantly less compact. Hence, the overall expressions for $\beta_{h_{abc}}$ and $\beta_{m^2_{ab}}$ are rather long compared to eq.~\eq{quartic-sum}ff. and are omitted. For the sake of reducing the cross section of transcription errors alone, the author recommends to automate the application of the dummy field method onto the latter equations instead.

\section{Example: Litim-Sannino Model}\label{sec:lisa}
In this section, we apply the obtained results by computing the three-loop RGEs of the Litim-Sannino model \cite{Litim:2014uca,Litim:2015iea} with vanishing gauge couplings. 
The model is described in terms of the Lagrangian
\begin{equation}\label{eq:LiSa-action}
\begin{aligned}
    \mathcal{L} = & \phantom{+}  \tr\left[\overline{\Psi}\,i\slashed{\partial}\,\Psi\right] + \tr\left[\partial_\mu H^\dagger\partial^\mu H\right] - m^2 \,\tr\left[H^\dagger H\right]  \\
    & - y\,\tr\left[\overline{\Psi} \, H P_R \,\Psi + \overline{\Psi}\, H^\dagger P_L \,\Psi \right] - u\,\tr\left[H^\dagger H H^\dagger H\right] - v \, \tr\left[H^\dagger H\right] \tr\left[H^\dagger H\right]
\end{aligned}
\end{equation}
and exhibits a $U(N_c) \times U(N_f)_L \times U(N_f)_R$ global symmetry. The Dirac fermions $\Psi$ and complex scalars $H$ are $N_c \times N_f$ and $N_f \times N_f$ matrix fields, respectively. $P_{L,R} = \tfrac12(1 \mp \gamma_5)$ in \eq{LiSa-action} represents the left- and right chiral projectors. 
Moreover, we employ the limit $N_{f,c} \rightarrow \infty$, while the quantity
\begin{equation}
  \epsilon = \frac{N_f}{N_c} - \frac{11}{2}
\end{equation}
remains finite, along with the rescaled couplings
\begin{equation}
  \alpha_y = \frac{N_c \, y^2}{(4 \pi)^2}, \qquad \alpha_u = \frac{N_f \,u}{(4 \pi)^2}, \qquad \alpha_v = \frac{N_f^2 \,v}{(4 \pi)^2}.
\end{equation}
In this limit, two-loop results are available from \cite{Bond:2017tbw}. At three-loop, all unknown coefficients drop out and one obtains
\begin{equation}
  \begin{aligned}
       \gamma_{H}\big|_\text{3-loop} = & - 4\, \alpha_u^3  - \tfrac{15}2 \alpha_u^2 \alpha_y +\left[ 5 \,\alpha_u  + \tfrac1{32}\left(183 + 10\epsilon\right) \alpha_y \right]\left(\tfrac{11}2 + \epsilon\right) \alpha_y^2, \\
     \gamma_{\Psi}\big|_\text{3-loop} = & \phantom{+} \left[ -\tfrac{11}{4} 
     \alpha_u^2  +\left(11 + 2\epsilon\right) \alpha_u \alpha_y +  \left(\tfrac{1217}{128} + \tfrac{41}{32} \epsilon - \tfrac{3}{32} \epsilon^2\right) \alpha_y^2 \right] \left(\tfrac{11}2 + \epsilon\right) \alpha_y , \\
     \beta_{\alpha_y}\big|_\text{3-loop} = & - 8\,\alpha_u^3 \alpha_y +  5\left(\tfrac52 + \epsilon\right) \alpha_u^2 \alpha_y^2 +  12 \left(\tfrac{11}2 + \epsilon\right)(8 + \epsilon)  \alpha_u \alpha_y^3  \\
     & +  \left(\tfrac{11}2 + \epsilon\right) \left(\tfrac{1583}{32} +\tfrac{23}{4} \epsilon  - \tfrac38 \epsilon^2\right) \alpha_y^4,\\
    \beta_{\alpha_u}\big|_\text{3-loop} = & \phantom{+} 104\, \alpha_u^4 + 34\, \alpha_u^3 \alpha_y  + \left(889 + 166 \epsilon\right) \alpha_u^2 \alpha_y^2  - \tfrac1{8}\left(\tfrac{11}2 + \epsilon\right)^2 (21 - 26 \epsilon) \alpha_y^4 \\
    & - \left(\tfrac{2953}{16} + \tfrac{315}8 \epsilon \right) \left(11 + 2\epsilon\right)\alpha_u \alpha_y^3 , \\
     \beta_{\alpha_v}\big|_\text{3-loop} = & \phantom{+} 12\, \alpha_v^2 \alpha_u^2 + 480\, \alpha_v \alpha_u^3 + (772 + 384 \zeta_3 ) \alpha_u^4  + 66\,\alpha_v \alpha_u^2 \alpha_y + 192\,\alpha_u^3 \alpha_y \\ 
     & + \left(\tfrac{427}2 + 41 \epsilon\right) \alpha_v^2 \alpha_y^2 + \left(788 + 152 \epsilon + 96\zeta_3 \left(\tfrac{11}2 + \epsilon\right)\right) \alpha_v \alpha_u \alpha_y^2 \\
     & + \left(\tfrac{1985}2 + 187 \epsilon + 192\zeta_3 \left(\tfrac{11}2 + \epsilon\right)\right) \alpha_u^2 \alpha_y^2 \\
     & - 4 \left(\tfrac{11}2 + \epsilon\right) \left(105 + 22 \epsilon + 24 \zeta_3 \left(\tfrac{11}2 + \epsilon\right)\right) \alpha_u \alpha_y^3\\
     & - \tfrac1{8} \left(\tfrac{11}2 + \epsilon\right)\left(1545  + 374 \epsilon\right) \alpha_v \alpha_y^3  - \left(\frac{11}2 + \epsilon\right)^2 \left(73 + 10 \epsilon \right) \alpha_y^4, \\
     m^{-2} \beta_{m^2}\big|_\text{3-loop} = &  \phantom{+} 12\,\alpha_v  \alpha_u^2 + 240 \alpha_u^3 + 33 \, \alpha_u^2 \alpha_y  + \left(394 + 76 \epsilon + 48 \zeta_3 \left(\tfrac{11}2 + \epsilon\right)\right) \alpha_u \alpha_y^2 \\
     & + (\tfrac{427}2 + 41 \epsilon) \alpha_v \alpha_y^2 - \tfrac1{16}\left(\tfrac{11}2 + \epsilon\right)\left(1545 + 374 \epsilon\right) \alpha_y^3 \\
  \end{aligned}
\end{equation}
as a consequence of \eq{gamma_s}, \eq{gamma_f}, \eq{beta_yuk} and \eq{quartic-sum}ff. These results are compatible with \cite{Steudtner:2020tzo,Bednyakov:2021clp}.

\section{Discussion and Outlook}\label{sec:dis}
In this work, we have investigated the three-loop renormalisation group evolution of general QFTs with scalars and fermions, featuring Yukawa, scalar quartic and cubic interactions, as well as fermion and scalar mass terms.
Using literature results, we have completely determined the general result for all field anomalous dimensions, Yukawa and fermion mass $\beta$-functions, 
as well as terms $\propto \lambda^4$, $\propto y^2 \lambda^3$ and $\propto y^4 \lambda^2$ for scalar quartics.
In the scalar potential, the number of unknown parameters has been vastly reduced to \openparams, which can be fixed by future calculations. That is, either by direct computation of the diagrams listed herein, or as a side product of determining RGEs in specific theories. 
 Unfortunately, there is little hope to utilise Weyl consistency conditions \cite{ Osborn:1991gm,Jack:2013sha,Poole:2019kcm} to constrain these coefficients, as five- and four-loop gauge and Yukawa $\beta$-functions would be required. Rather, it is the other way around that computations of the latter profit from three-loop quartic RGEs. 
The results obtained here may already be sufficient to completely determine three-loop RGEs for certain QFTs. Trivial examples include the works \cite{Chetyrkin:2012rz,Bednyakov:2013eba,Chetyrkin:2013wya,Bednyakov:2013cpa,Bednyakov:2014pia,Zerf:2017zqi,Mihaila:2017ble} used as input, and it was demonstrated in Sec.~\ref{sec:lisa} that this selection is not exhaustive.
In the absence of fermion masses, only three unknown parameters remain in the general RG evolution of scalar mass and cubic interactions.

Once the remaining parameters in this setup are determined, the next step towards a complete three-loop result would be the inclusion of gauge interactions. The gauge $\beta$-function itself is already known at three-loop \cite{Pickering:2001aq, Mihaila:2012pz, Mihaila:2014caa, Poole:2019kcm}, and progress towards a general four-loop expression is being made \cite{Bednyakov:2015ooa,Zoller:2015tha,Poole:2019txl, Poole:2019kcm,Davies:2019onf}. In particular, a basis of tensor structures has been formulated in \cite{Poole:2019kcm}, which also covers three-loop fermion and scalar anomalous dimensions, as well as Yukawa $\beta$-functions. The work spells out explicitly how involved the determination of these fully general RGEs is -- an effort likely to be eclipsed by the three-loop quartic $\beta$-functions, for which a complete basis remains to be determined. This vision profits from the gaugeless contribution to be handled separately here in this work.
As for the coefficients involving gauge interactions in $\gamma^{\phi,3\ell}$, $\gamma^{\psi,3\ell}$ and $\beta_{y^a}^{3\ell}$, substantial amount may be fixed using the explicit computation \cite{Herren:2017uxn} and the conditions in \cite{Poole:2019kcm}. Likewise, as the SM gauge group is relatively large and structurally diverse, one should be cautiously optimistic that available results \cite{Chetyrkin:2012rz,Bednyakov:2013eba,Chetyrkin:2013wya,Bednyakov:2013cpa,Bednyakov:2014pia} determine a sizeable number of  contributions to $\beta_{\lambda_{abcd}}^{3\ell}$.

\section*{Acknowledgements}
The author is thankful to Ian Jack and Hugh Osborn for comments on the manuscript as well as various insightful discussions, Dominik Stöckinger and  Sandra Kvedarait\.e for useful comments as well as Gustavo Medina Vazquez for proof-reading.

\appendix

\section{Conversion to Weyl Consistency Condition Basis}\label{app:WeylCC}
In this appendix, we provide the results \eq{beta_yuk} in the basis of \cite{Poole:2019kcm}, wherein Weyl consistency condition between four-loop gauge, three-loop Yukawa and two-loop quartic RGEs have been studied. With the help of the program \cite{Thomsen:2021ncy}, we obtain the 86 coefficients
\begin{equation}\label{eq:aet}
  \begin{array}{llll}
    \ccy{43\phantom{0}} = -\tfrac1{16} , & 
    \ccy{128} = \tfrac32 , \qquad& 
    \ccy{129} = -\tfrac38 , \qquad&
    \ccy{130} = -\tfrac{11}{48} ,  \\[.5em]
    \ccy{131} = 1 ,&
    \ccy{132} = -\tfrac5{32} , &
    \ccy{224} = 3 , &
    \ccy{225} = -2 , \\[.5em]
    \ccy{226} = 4 , &
    \ccy{227} = 5 , &
    \ccy{228} = 2 , &
    \ccy{229} = 6 , \\[.5em]
    \ccy{230} = 2 , &
    \ccy{231} = \tfrac58 , &
    \ccy{232} = \tfrac58 , &
    \ccy{233} = 1 , \\[.5em]
    \ccy{234} = \tfrac32 , &
    \ccy{235} = 1 , &
    \ccy{236} = -3 , &
    \ccy{237} = 4(3 \zeta_3 - 1) , \\[.5em]
    \ccy{238} = 6(2 \zeta_3 - 1) ,\quad &
    \ccy{239} = 4(3 \zeta_3 - 2) , \quad&
    \ccy{240} = 2 , &
    \ccy{241} = 2 , \\[.5em]
    \ccy{242} = 4(3 \zeta_3 - 1) ,\quad &
    \ccy{243} = 6(2 \zeta_3 - 1) , \quad&
    \ccy{244} = 6 \zeta_3 - 5 ,\quad &
    \ccy{245} = 3 \zeta_3 - 2 , \\[.5em]
    \ccy{246} = \tfrac32 \zeta_3 - 1 , &
    \ccy{247} = -3 , &
    \ccy{248} = -1 , &
    \ccy{249} = -1 , \\[.5em]
    \ccy{250} = -8 , &
    \ccy{251} = -6 , &
    \ccy{252} = 0 , &
    \ccy{253} = -2 ,\\[.5em]
    \ccy{254} = -4 , &
    \ccy{255} = 2 , &
    \ccy{256} = 4 , &
    \ccy{257} = \tfrac54 ,\\[.5em]
    \ccy{258} = -2 , &
    \ccy{259} = 4 , &
    \ccy{260} = -2 , &
    \ccy{261} = -\tfrac58 ,\\[.5em]
    \ccy{262} = 1 , &
    \ccy{263} = 1 , &
    \ccy{264} = 0 , &
    \ccy{265} = 0 ,\\[.5em]
    \ccy{266} = -\tfrac38 , &
    \ccy{267} = -\tfrac34 , &
    \ccy{268} = \tfrac74 , &
    \ccy{269} = 1 ,\\[.5em]
    \ccy{270} = \tfrac{25}8 , &
    \ccy{271} = \tfrac78 , &
    \ccy{272} = -3 , &
    \ccy{273} = 4 ,\\[.5em]
    \ccy{274} = 3 , &
    \ccy{275} = -2 , &
    \ccy{276} = -3 , &
    \ccy{277} = \tfrac3{16} ,\\[.5em]
    \ccy{278} = \tfrac12 , &
    \ccy{279} = 2 , &
    \ccy{280} = \tfrac18 , &
    \ccy{281} = \tfrac3{16} ,\\[.5em]
    \ccy{282} = \tfrac7{16} , &
    \ccy{283} = \tfrac5{16} , &
    \ccy{284} = \tfrac7{16} , &
    \ccy{285} = 2 ,\\[.5em]
    \ccy{286} = \tfrac{25}{8} , &
    \ccy{287} = -3 , &
    \ccy{288} = 3 , &
    \ccy{289} = -1 ,\\[.5em]
    \ccy{290} = -\tfrac{3}{16} , &
    \ccy{291} = -\tfrac9{16} , &
    \ccy{292} = -\tfrac3{16} , &
    \ccy{293} = \tfrac9{16} ,\\[.5em]
    \ccy{294} = 1 , &
    \ccy{295} = -\tfrac12 , &
    \ccy{296} = -1 , &
    \ccy{297} = -\tfrac5{16} ,\\[.5em]
    \ccy{298} = \tfrac1{32} , &
    \ccy{299} = -\tfrac3{16} , &
    \ccy{300} = -1 , &
    \ccy{301} = -\tfrac1{4} ,\\[.5em]
    \ccy{302} = -\tfrac1{2} , &
    \ccy{303} = -\tfrac3{16} .
  \end{array}
\end{equation}
Our findings are not in contradiction to the general 265 Weyl consistency condition formulated in \cite{Poole:2019kcm}. In fact, $\ccy{226}$ and $\ccy{228}$ fit their direct prediction therein. While the two-loop quartic RGEs merely reduce the number of conditions by 4, the results \eq{aet} leave only 219 such conditions. In particular, the direct predictions of coefficients are updated to
\begin{equation}
  \begin{array}{lllll}
  \ccg{55\phantom{0}} = -\tfrac{27}2 ,\quad &
  \ccg{60\phantom{0}} = \tfrac12 , \quad&
  \ccg{67\phantom{0}} = -\tfrac1{12} , \quad&
  \ccg{110} = -4  ,\quad &
  \ccg{112} = \tfrac14 , \\[.5em]
  \ccg{158} = 0 , &
  \ccg{160} = -\tfrac52 , &
  \ccy{2\phantom{00}} = -3 , \quad&
  \ccy{4\phantom{00}} = -\tfrac32 , \quad&
  \ccy{39\phantom{0}}  = -12 , \\[.5em]
  \ccy{40\phantom{0}} = \tfrac54 , \quad&
  \ccy{41\phantom{0}} = -\tfrac{17}{48} , \quad&
  \ccy{42\phantom{0}} = \tfrac{19}{16} , \quad&
  \ccy{47\phantom{0}} = \tfrac34 , \quad&
  \ccy{49\phantom{0}}  = 0 , \\[.5em]
  \ccy{122} = 0 , \quad&
  \ccy{123} = -\tfrac{17}{2} , \quad&
  \ccy{124} = -17 , \quad&
  \ccy{125} = \tfrac{19}2 , \quad&
  \ccy{126}  = 2 , \\[.5em]
  \ccy{127} = 0 .
  \end{array}
\end{equation}

\section{Comparison to Supersymmetric RGEs}\label{app:susy}

\begin{table}
  \centering
  \renewcommand{\arraystretch}{1}
  \begin{tabular}{|c|l|}
    \hline
    \cell{c}{\begin{tikzpicture}
    \draw[black, thick] (0.em, 1.em) -- (1em,0em) -- (2em,0) --(3em,1em) -- (4em,1em);
     \draw[black, thick] (0.em, -1.em) -- (1em,0em);
     \draw[black, thick] (2em,0) --(3em,-1em) -- (4em,-1em);
     \draw[black, thick] (4.5em,1em) circle (.5em);
     \draw[black, thick] (4.5em,-1em) circle (.5em);
     \draw[black, thick] (5em,1em) -- (6em,1em) -- (7em,0em) -- (8em,0em) -- (9em,1em);
     \draw[black, thick] (5em,-1em) -- (6em,-1em)  -- (7em,0em) -- (8em,0em) -- (9em,-1em);
     \node at (1em, 0em)[circle,fill,inner sep=.1em]{};
     \node at (2em, 0em)[fill,inner sep=.125em]{};
     \node at (3em, 1em)[circle,fill,inner sep=.1em]{};
     \node at (3em, -1em)[circle,fill,inner sep=.1em]{};
     \node at (4em, 1em)[fill,inner sep=.125em]{};
     \node at (4em, -1em)[fill,inner sep=.125em]{};
     \node at (5em, 1em)[circle,fill,inner sep=.1em]{};
     \node at (5em, -1em)[circle,fill,inner sep=.1em]{};
     \node at (6em, 1em)[fill,inner sep=.125em]{};
     \node at (6em, -1em)[fill,inner sep=.125em]{};
     \node at (7em, 0em)[circle,fill,inner sep=.1em]{};
     \node at (8em, 0em)[fill,inner sep=.125em]{};
    \end{tikzpicture}}
    & $c_{(3)} = - \tfrac14$ \\
    \cell{c}{\begin{tikzpicture}
     \draw[black, thick] (0.em, 1.em) -- (1em,0em) -- (2em,0) --(3em,1em) -- (3.5em,1em);
     \draw[black, thick] (4.5em,1em) -- (5em,1em);
     \draw[black, thick] (6em,1em) -- (6.5em,1em) -- (7.5em, 0em) -- (8.5em,0) -- (9.5em,1em);
     \draw[black, thick] (8.5em,0) -- (9.5em,-1em);
     \draw[black, thick] (0.em, -1.em) -- (1em,0em);
     \draw[black, thick] (2em,0) --(3em,-1em) -- (6.5em,-1em) -- (7.5em, 0);
     \draw[black, thick] (4em,1em) circle (.5em);
     \draw[black, thick] (5.5em,1em) circle (.5em);
     \node at (1em, 0em)[circle,fill,inner sep=.1em]{};
     \node at (2em, 0em)[fill,inner sep=.125em]{};
     \node at (3em, 1em)[circle,fill,inner sep=.1em]{};
     \node at (3em, -1em)[circle,fill,inner sep=.1em]{};
     \node at (3.5em, 1em)[fill,inner sep=.125em]{};
     \node at (4.5em, 1em)[circle,fill,inner sep=.1em]{};
     \node at (5em, 1em)[fill,inner sep=.125em]{};
     \node at (6em, 1em)[circle,fill,inner sep=.1em]{};
     \node at (6.5em, 1em)[fill,inner sep=.125em]{};
     \node at (6.5em, -1em)[fill,inner sep=.125em]{};
     \node at (7.5em, 0em)[circle,fill,inner sep=.1em]{};
     \node at (8.5em, 0em)[fill,inner sep=.125em]{};
    \end{tikzpicture}}
    & $c_{(4)} = - \tfrac12$\\
    \cell{c}{\begin{tikzpicture}
    \draw[black, thick] (0.em, -1.em) -- (1em,0em);
     \draw[black, thick] (0.em, 1.em) -- (1em,0em) -- (2em,0) --(3em,1em) -- (3.5em,1em) -- (4.5em,0em) -- (5em,0em);
     \draw[black, thick] (5.5em, 0em) circle (.5em);
     \draw[black, thick] (6em,0em) -- (6.5em,0em) -- (7.5em,1em) -- (8em,1em) -- (9em, 0em) -- (10em, 0em) -- (11em, 1em);
     \draw[black, thick] (10em, 0em) -- (11em, -1em); 
     \draw[black, thick] (2em,0) -- (3em,-1em) -- (8em,-1em) -- (9em, 0em);
     \draw[black, thick] (3.5em,1em) -- (8em,1em);
     \node at (1em, 0em)[circle,fill,inner sep=.1em]{};
     \node at (2em, 0em)[fill,inner sep=.125em]{};
     \node at (3em, 1em)[circle,fill,inner sep=.1em]{};
     \node at (3em, -1em)[circle,fill,inner sep=.1em]{};
     \node at (3.5em, 1em)[fill,inner sep=.125em]{};
     \node at (4.5em, 0em)[circle,fill,inner sep=.1em]{};
     \node at (5em, 0em)[fill,inner sep=.125em]{};
     \node at (6em, 0em)[circle,fill,inner sep=.1em]{};
     \node at (6.5em, 0em)[fill,inner sep=.125em]{};
     \node at (7.5em, 1em)[circle,fill,inner sep=.1em]{};
     \node at (8em, 1em)[fill,inner sep=.125em]{};
     \node at (8em, -1em)[fill,inner sep=.125em]{};
     \node at (9em, 0em)[circle,fill,inner sep=.1em]{};
     \node at (10em, 0em)[fill,inner sep=.125em]{};
    \end{tikzpicture}}
    & $c_{(6)} = \tfrac{25}8$\\
     \cell{c}{\begin{tikzpicture}
    \draw[black, thick] (0.em, -1.em) -- (1em,0em);
     \draw[black, thick] (0.em, 1.em) -- (1em,0em) -- (2em,0) --(3em,0em) -- (4em,0em) -- (5em,0em) -- (6em,0em) -- (7em,0em) -- (8em,0em) -- (9em,1em);
      \draw[black, thick] (8em,0em) -- (9em,-1em);
      \draw[black, thick] (2em,0em) to [bend left=60] (5em, 0em);
      \draw[black, thick] (3em,0em) to [bend right=60] (6em, 0em);
      \draw[black, thick] (4em,0em) to [bend left=60] (7em, 0em);
     \node at (1em, 0em)[circle,fill,inner sep=.1em]{};
     \node at (2em, 0em)[fill,inner sep=.125em]{};
     \node at (3em, 0em)[circle,fill,inner sep=.1em]{};
     \node at (4em, 0em)[fill,inner sep=.125em]{};
     \node at (5em, 0em)[circle,fill,inner sep=.1em]{};
     \node at (6em, 0em)[fill,inner sep=.125em]{};
     \node at (7em, 0em)[circle,fill,inner sep=.1em]{};
     \node at (8em, 0em)[fill,inner sep=.125em]{};
    \end{tikzpicture}}
    & $c_{(8)} = 3\zeta_3-1$\\
    \cell{c}{\begin{tikzpicture}
    \draw[black, thick] (0.em, -1.em) -- (1em,0em);
     \draw[black, thick] (0.em, 1.em) -- (1em,0em) -- (2em,0) --(3em,0em) -- (4em,0em) -- (5em,0em) -- (6em,0em) -- (7em,0em) -- (8em,0em) -- (9em,0em);
      \draw[black, thick] (2em,0em) to [bend left=50] (7em, 0em);
      \draw[black, thick] (3em,0em) to [bend left=50] (6em, 0em);
      \draw[black, thick] (4em,0em) to (4em, -.5em);
      \draw[black, thick] (5em,0em) to [bend right=60] (8em, 0em);
     \node at (1em, 0em)[circle,fill,inner sep=.1em]{};
     \node at (2em, 0em)[fill,inner sep=.125em]{};
     \node at (3em, 0em)[circle,fill,inner sep=.1em]{};
     \node at (4em, 0em)[fill,inner sep=.125em]{};
     \node at (5em, 0em)[circle,fill,inner sep=.1em]{};
     \node at (6em, 0em)[fill,inner sep=.125em]{};
     \node at (7em, 0em)[circle,fill,inner sep=.1em]{};
     \node at (8em, 0em)[fill,inner sep=.125em]{};
    \end{tikzpicture}}
    & $c_{(14)} = -6 - 2\,c_{(11)}$\\
    \cell{c}{\begin{tikzpicture}
    \draw[black, thick] (0.em, -1.em) -- (1em,0em);
     \draw[black, thick] (0.em, 1.em) -- (1em,0em) -- (2em,0) --(3em,0em) ;
     \draw[black, thick]  (4em,0em) -- (5em,0em) -- (6em,0em) -- (7em,0em) -- (8em,0em) -- (9em,0em);
      \draw[black, thick] (2em,0em) to [bend left=40] (7em, 0em);
      \draw[black, thick] (3.5em,0em) circle (.5em);
      \draw[black, thick] (6em,0em) to (6em, -.5em);
      \draw[black, thick] (5em,0em) to [bend left=60] (8em, 0em);
     \node at (1em, 0em)[circle,fill,inner sep=.1em]{};
     \node at (2em, 0em)[fill,inner sep=.125em]{};
     \node at (3em, 0em)[circle,fill,inner sep=.1em]{};
     \node at (4em, 0em)[fill,inner sep=.125em]{};
     \node at (5em, 0em)[circle,fill,inner sep=.1em]{};
     \node at (6em, 0em)[fill,inner sep=.125em]{};
     \node at (7em, 0em)[circle,fill,inner sep=.1em]{};
     \node at (8em, 0em)[fill,inner sep=.125em]{};
    \end{tikzpicture}}
    & $c_{(21)} = -4\,c_{(2)}$\\
    \cell{c}{\begin{tikzpicture}
    \draw[black, thick] (0.em, -1.em) -- (1em,0em);
     \draw[black, thick] (0.em, 1.em) -- (1em,0em) -- (2em,0) --(3em,0em) -- (4em,0em) -- (5em,0em) -- (6em,0em) ; 
     \draw[black, thick] (7em,0em) -- (8em,0em) -- (9em,0em);
      \draw[black, thick] (2em,0em) to [bend left=60] (5em, 0em);
      \draw[black, thick] (3em,0em) to [bend left=40] (8em, 0em);
      \draw[black, thick] (4em,0em) to (4em, -.5em);
      \draw[black, thick] (6.5em,0em) circle (.5em);
     \node at (1em, 0em)[circle,fill,inner sep=.1em]{};
     \node at (2em, 0em)[fill,inner sep=.125em]{};
     \node at (3em, 0em)[circle,fill,inner sep=.1em]{};
     \node at (4em, 0em)[fill,inner sep=.125em]{};
     \node at (5em, 0em)[circle,fill,inner sep=.1em]{};
     \node at (6em, 0em)[fill,inner sep=.125em]{};
     \node at (7em, 0em)[circle,fill,inner sep=.1em]{};
     \node at (8em, 0em)[fill,inner sep=.125em]{};
    \end{tikzpicture}}
    & $c_{(25)} = -2 - c_{(1)}$\\
    \cell{c}{\begin{tikzpicture}
     \draw[black, thick] (0.em, 0em) -- (1em,0em) -- (2em,0) --(3em,0em) -- (4em,0em) -- (5em,0em) -- (6em,0em) -- (7em,0em) -- (8em,0em);
      \draw[black, thick] (1em,0em) to [bend right=25] (8em, 0em);
      \draw[black, thick] (2em,0em) to (2em, .5em);
      \draw[black, thick] (5em,0em) to (5em, -.5em);
      \draw[black, thick] (6em,0em) to (6em, -.5em);
      \draw[black, thick] (3em,0em) to [bend left=45] (8em, 0em);
      \draw[black, thick] (4em,0em) to [bend left=45] (7em, 0em);
     \node at (1em, 0em)[circle,fill,inner sep=.1em]{};
     \node at (2em, 0em)[fill,inner sep=.125em]{};
     \node at (3em, 0em)[circle,fill,inner sep=.1em]{};
     \node at (4em, 0em)[fill,inner sep=.125em]{};
     \node at (5em, 0em)[circle,fill,inner sep=.1em]{};
     \node at (6em, 0em)[fill,inner sep=.125em]{};
     \node at (7em, 0em)[circle,fill,inner sep=.1em]{};
     \node at (8em, 0em)[fill,inner sep=.125em]{};
    \end{tikzpicture}}
    & $c_{(29)} = -2\,c_{(7)} -4 \,c_{(9)} -2\,c_{(13)}  - c_{(47)}$\\
    \cell{c}{\begin{tikzpicture}
     \draw[black, thick] (0.em, 0em) -- (1em,0em) -- (2em,0) --(3em,0em) -- (4em,0em) -- (5em,0em) -- (6em,0em) -- (7em,0em) -- (8em,0em) -- (9em,0em);
      \draw[black, thick] (1em,0em) to [bend right=60] (4em, 0em);
      \draw[black, thick] (2em,0em)  to [bend left=45] (7em, 0em);
      \draw[black, thick] (3em,0em) to (3em, .5em);
      \draw[black, thick] (5em,0em) to [bend right=60] (8em, 0em);
      \draw[black, thick] (6em,0em) to (6em, .5em);
     \node at (1em, 0em)[circle,fill,inner sep=.1em]{};
     \node at (2em, 0em)[fill,inner sep=.125em]{};
     \node at (3em, 0em)[circle,fill,inner sep=.1em]{};
     \node at (4em, 0em)[fill,inner sep=.125em]{};
     \node at (5em, 0em)[circle,fill,inner sep=.1em]{};
     \node at (6em, 0em)[fill,inner sep=.125em]{};
     \node at (7em, 0em)[circle,fill,inner sep=.1em]{};
     \node at (8em, 0em)[fill,inner sep=.125em]{};
    \end{tikzpicture}}
    & $c_{(31)} = 4 - 8\,c_{(10)}$\\
    \cell{c}{\begin{tikzpicture}
     \draw[black, thick] (0.em, 0em) -- (1em,0em) -- (2em,0) --(3em,0em) -- (4em,0em) -- (5em,0em) -- (6em,0em) -- (7em,0em) -- (8em,0em) -- (9em,0em);
      \draw[black, thick] (1em,0em) to [bend right=30] (6em, 0em);
      \draw[black, thick] (2em,0em)  to (2em,.5em) ;
      \draw[black, thick] (3em,0em) to [bend right=30] (8em, 0em);
      \draw[black, thick] (4em,0em) arc (180:0:.5em);
      \draw[black, thick] (7em,0em)  to (7em,.5em) ;
     \node at (1em, 0em)[circle,fill,inner sep=.1em]{};
     \node at (2em, 0em)[fill,inner sep=.125em]{};
     \node at (3em, 0em)[circle,fill,inner sep=.1em]{};
     \node at (4em, 0em)[fill,inner sep=.125em]{};
     \node at (5em, 0em)[circle,fill,inner sep=.1em]{};
     \node at (6em, 0em)[fill,inner sep=.125em]{};
     \node at (7em, 0em)[circle,fill,inner sep=.1em]{};
     \node at (8em, 0em)[fill,inner sep=.125em]{};
    \end{tikzpicture}}
    & $c_{(32)} = -1 - 2\,c_{(1)}- 2\,c_{(16)}$\\
    \cell{c}{\begin{tikzpicture}
     \draw[black, thick] (0.em, 0em) -- (1em,0em) -- (2em,0) --(3em,0em) -- (4em,0em) -- (5em,0em) -- (6em,0em) -- (7em,0em) -- (8em,0em);
      \draw[black, thick] (1em,0em) to [bend right=20] (8em, 0em);
      \draw[black, thick] (2em,0em) to (2em, .5em);
      \draw[black, thick] (3em,0em) to (3em, .5em);
      \draw[black, thick] (4em,0em) to (4em, .5em);
      \draw[black, thick] (5em,0em) to [bend left=60] (8em, 0em);
      \draw[black, thick] (6em,0em) arc (180:0:.5em);
     \node at (1em, 0em)[circle,fill,inner sep=.1em]{};
     \node at (2em, 0em)[fill,inner sep=.125em]{};
     \node at (3em, 0em)[circle,fill,inner sep=.1em]{};
     \node at (4em, 0em)[fill,inner sep=.125em]{};
     \node at (5em, 0em)[circle,fill,inner sep=.1em]{};
     \node at (6em, 0em)[fill,inner sep=.125em]{};
     \node at (7em, 0em)[circle,fill,inner sep=.1em]{};
     \node at (8em, 0em)[fill,inner sep=.125em]{};
    \end{tikzpicture}}
    & $c_{(34)} = -2 - c_{(1)} - c_{(36)}$\\
        \cell{c}{\begin{tikzpicture}
     \draw[black, thick] (0.em, 0em) -- (1em,0em) -- (2em,0) --(3em,0em) -- (4em,0em) -- (5em,0em) -- (6em,0em) -- (7em,0em) -- (8em,0em);
      \draw[black, thick] (1em,0em) to [bend right=20] (8em, 0em);
      \draw[black, thick] (2em,0em) to (2em, .5em);
      \draw[black, thick] (3em,0em) to (3em, .5em);
      \draw[black, thick] (4em,0em) to (4em, .5em);
      \draw[black, thick] (5em,0em) arc(180:0:.5em);
      \draw[black, thick] (7em,0em) arc(180:0:.5em);
     \node at (1em, 0em)[circle,fill,inner sep=.1em]{};
     \node at (2em, 0em)[fill,inner sep=.125em]{};
     \node at (3em, 0em)[circle,fill,inner sep=.1em]{};
     \node at (4em, 0em)[fill,inner sep=.125em]{};
     \node at (5em, 0em)[circle,fill,inner sep=.1em]{};
     \node at (6em, 0em)[fill,inner sep=.125em]{};
     \node at (7em, 0em)[circle,fill,inner sep=.1em]{};
     \node at (8em, 0em)[fill,inner sep=.125em]{};
    \end{tikzpicture}}
    & $c_{(35)} = 1$\\
    \cell{c}{\begin{tikzpicture}
     \draw[black, thick] (0.em, 0em) -- (1em,0em) -- (2em,0) --(3em,0em) -- (4em,0em) -- (5em,0em) -- (6em,0em) -- (7em,0em) -- (8em,0em);
      \draw[black, thick] (1em,0em) to [bend right=30] (8em, 0em);
      \draw[black, thick] (2em,0em) to (2em, .5em);
      \draw[black, thick] (3em,0em) to [bend right=60] (6em, 0em);
      \draw[black, thick] (4em,0em) to (4em, .5em);
      \draw[black, thick] (5em,0em) to (5em, .5em);
      \draw[black, thick] (7em,0em) arc (180:0:.5em);
     \node at (1em, 0em)[circle,fill,inner sep=.1em]{};
     \node at (2em, 0em)[fill,inner sep=.125em]{};
     \node at (3em, 0em)[circle,fill,inner sep=.1em]{};
     \node at (4em, 0em)[fill,inner sep=.125em]{};
     \node at (5em, 0em)[circle,fill,inner sep=.1em]{};
     \node at (6em, 0em)[fill,inner sep=.125em]{};
     \node at (7em, 0em)[circle,fill,inner sep=.1em]{};
     \node at (8em, 0em)[fill,inner sep=.125em]{};
    \end{tikzpicture}}
    & $c_{(41)} = -2 - 2\,c_{(2)} - c_{(17)}- c_{(20)}$\\
    \cell{c}{\begin{tikzpicture}
     \draw[black, thick] (0.em, 0em) -- (1em,0em) -- (2em,0) --(3em,0em) -- (4em,0em) -- (5em,0em) -- (6em,0em) -- (7em,0em) -- (8em,0em);
      \draw[black, thick] (1em,0em) to [bend right=20] (8em, 0em);
      \draw[black, thick] (2em,0em) to (2em, .5em);
      \draw[black, thick] (3em,0em) to (3em, .5em);
      \draw[black, thick] (6em,0em) to (6em, .5em);
      \draw[black, thick] (4em,0em) arc(180:0:.5em);
      \draw[black, thick] (7em,0em) arc(180:0:.5em);
     \node at (1em, 0em)[circle,fill,inner sep=.1em]{};
     \node at (2em, 0em)[fill,inner sep=.125em]{};
     \node at (3em, 0em)[circle,fill,inner sep=.1em]{};
     \node at (4em, 0em)[fill,inner sep=.125em]{};
     \node at (5em, 0em)[circle,fill,inner sep=.1em]{};
     \node at (6em, 0em)[fill,inner sep=.125em]{};
     \node at (7em, 0em)[circle,fill,inner sep=.1em]{};
     \node at (8em, 0em)[fill,inner sep=.125em]{};
    \end{tikzpicture}}
    & $c_{(43)} = - 2\,c_{(2)}$\\
        \cell{c}{\begin{tikzpicture}
     \draw[black, thick] (0.em, 0em) -- (1em,0em) -- (2em,0) --(3em,0em) -- (4em,0em) -- (5em,0em) -- (6em,0em) -- (7em,0em) -- (8em,0em);
      \draw[black, thick] (1em,0em) to [bend right=20] (8em, 0em);
      \draw[black, thick] (2em,0em) to (2em, .5em);
      \draw[black, thick] (3em,0em) to (3em, .5em);
      \draw[black, thick] (6em,0em) to (6em, -.5em);
      \draw[black, thick] (4em,0em) to [bend left=60] (7em, 0em);
      \draw[black, thick] (5em,0em) to [bend left=60] (8em, 0em);
     \node at (1em, 0em)[circle,fill,inner sep=.1em]{};
     \node at (2em, 0em)[fill,inner sep=.125em]{};
     \node at (3em, 0em)[circle,fill,inner sep=.1em]{};
     \node at (4em, 0em)[fill,inner sep=.125em]{};
     \node at (5em, 0em)[circle,fill,inner sep=.1em]{};
     \node at (6em, 0em)[fill,inner sep=.125em]{};
     \node at (7em, 0em)[circle,fill,inner sep=.1em]{};
     \node at (8em, 0em)[fill,inner sep=.125em]{};
    \end{tikzpicture}}
    & $c_{(45)} = -8 - 2\,c_{(12)} - c_{(26)}$\\
    \cell{c}{\begin{tikzpicture}
     \draw[black, thick] (0.em, 0em) -- (1em,0em) -- (2em,0) --(3em,0em) -- (4em,0em) -- (5em,0em) -- (6em,0em) -- (7em,0em) -- (8em,0em);
      \draw[black, thick] (1em,0em) to [bend right=20] (8em, 0em);
      \draw[black, thick] (2em,0em) to (2em, .5em);
      \draw[black, thick] (5em,0em) to (5em, .5em);
      \draw[black, thick] (6em,0em) to (6em, .5em);
      \draw[black, thick] (3em,0em) arc(180:0:.5em);
      \draw[black, thick] (7em,0em) arc(180:0:.5em);
     \node at (1em, 0em)[circle,fill,inner sep=.1em]{};
     \node at (2em, 0em)[fill,inner sep=.125em]{};
     \node at (3em, 0em)[circle,fill,inner sep=.1em]{};
     \node at (4em, 0em)[fill,inner sep=.125em]{};
     \node at (5em, 0em)[circle,fill,inner sep=.1em]{};
     \node at (6em, 0em)[fill,inner sep=.125em]{};
     \node at (7em, 0em)[circle,fill,inner sep=.1em]{};
     \node at (8em, 0em)[fill,inner sep=.125em]{};
    \end{tikzpicture}}
    & $c_{(46)} = 1$\\
    \cell{c}{\begin{tikzpicture}
     \draw[black, thick] (0.em, 0em) -- (1em,0em) -- (2em,0) --(3em,0em) -- (4em,0em) -- (5em,0em) -- (6em,0em) -- (7em,0em) -- (8em,0em);
      \draw[black, thick] (1em,0em) to [bend right=20] (8em, 0em);
      \draw[black, thick] (2em,0em) to (2em, .5em);
      \draw[black, thick] (3em,0em) to [bend left=40] (8em, 0em);
      \draw[black, thick] (4em,0em) to (4em, .5em);
      \draw[black, thick] (5em,0em) arc (180:0:.5em);
      \draw[black, thick] (7em,0em) to (7em, .5em);
     \node at (1em, 0em)[circle,fill,inner sep=.1em]{};
     \node at (2em, 0em)[fill,inner sep=.125em]{};
     \node at (3em, 0em)[circle,fill,inner sep=.1em]{};
     \node at (4em, 0em)[fill,inner sep=.125em]{};
     \node at (5em, 0em)[circle,fill,inner sep=.1em]{};
     \node at (6em, 0em)[fill,inner sep=.125em]{};
     \node at (7em, 0em)[circle,fill,inner sep=.1em]{};
     \node at (8em, 0em)[fill,inner sep=.125em]{};
    \end{tikzpicture}}
    & $c_{(52)} = 1- c_{(1)} - c_{(9)}- c_{(15)}$\\
    \cell{c}{\begin{tikzpicture}
     \draw[black, thick] (0.em, 0em) -- (1em,0em) -- (2em,0) --(3em,0em) -- (4em,0em) -- (5em,0em) -- (6em,0em) -- (7em,0em) -- (8em,0em);
      \draw[black, thick] (1em,0em) to [bend right=20] (8em, 0em);
      \draw[black, thick] (2em,0em) to (2em, .5em);
      \draw[black, thick] (3em,0em) to [bend left=60] (6em, 0em);
      \draw[black, thick] (4em,0em) to (4em, .5em);
      \draw[black, thick] (5em,0em) to [bend left=60] (8em, 0em);
      \draw[black, thick] (7em,0em) to (7em, .5em);
     \node at (1em, 0em)[circle,fill,inner sep=.1em]{};
     \node at (2em, 0em)[fill,inner sep=.125em]{};
     \node at (3em, 0em)[circle,fill,inner sep=.1em]{};
     \node at (4em, 0em)[fill,inner sep=.125em]{};
     \node at (5em, 0em)[circle,fill,inner sep=.1em]{};
     \node at (6em, 0em)[fill,inner sep=.125em]{};
     \node at (7em, 0em)[circle,fill,inner sep=.1em]{};
     \node at (8em, 0em)[fill,inner sep=.125em]{};
    \end{tikzpicture}}
    & $c_{(54)} = -6(1 + 2 \zeta_3) - 2\, c_{(11)} - c_{(22)}- 2\,c_{(28)}$\\
    \hline
  \end{tabular}
  \caption{Relations between coefficients for tensor structures in \fig{quart1} and \fig{quart2} (left), inferred from comparison with supersymmetric RGEs. The corresponding graphs on the right contain round vertices denoting an insertion of $Y^{ABC}$ while squared nodes mark $Y_{ABC}$ interactions. }
  \label{tab:susy-relations}
\end{table}

The non-renormalisation theorem \cite{Salam:1974jj,Grisaru:1979wc} suggests that only superfield anomalous dimensions $\gamma^A_{\phantom{A}B}$, contribute to the $\beta$-functions of superpotential parameters, while quantum corrections to the vertex renormalisations are finite. Expressions for the superfield anomalous dimensions are available at three-loop \cite{Jack:1996qq,Parkes:1985hh} or using dimensional reduction (DRED) \cite{Siegel:1979wq,Capper:1979ns} and the $\overline{\text{DR}}$ scheme.
In renormalisable QFTs without gauge interactions, DRED is equivalent to DREG since no $\epsilon$-scalars are introduced, up to ambiguities due to the $\gamma_5$ which are absent in our setup as argued earlier.\footnote{See \cite{Gnendiger:2017pys} for a recent overview of various regularisation schemes.} At two-loop order, this is verified explicitly by the scheme conversions provided by \cite{Martin:1993yx}.
  Therefore, we can express $\mathcal{N} = 1$ QFTs in the language of a non-supersymmetric ones, see for instance \cite{Martin:1993yx,Martin:1997ns}, and directly compare our $\overline{\text{MS}}$ ansatz against the supersymmetric RGEs in the $\overline{\text{DR}}$ scheme -- similar to the conduct in \cite{Jack:2013sha,Jack:2014pua,Schienbein:2018fsw}. 
The perturbatively renormalisable $\mathcal{N} = 1$ superpotential of dimensionless couplings
\begin{equation}
  W = \tfrac16 Y^{ABC}\Phi_A \Phi_B \Phi_C, 
\end{equation} 
only contains Yukawa interactions $Y^{ABC}$ and is holomorphic in the chiral superfields $\Phi_A$. Here, the corresponding complex conjugate Yukawa interaction will be denoted as $Y_{ABC}$, counting only over antichiral fields $\Phi^A$. Each superfield $\Phi_A$ contains the complex scalar $\phi_A$ and Weyl fermion $\psi_A$ as well as auxiliary fields. Integrating the latter out from the action, one obtains the Yukawa and quartic interactions 
\begin{equation}
  \tfrac12 Y^{ABC} \left(\phi_A^{\phantom{\intercal}}  \psi_B^\intercal \varepsilon \psi_C^{\phantom{\intercal}} \right) + \text{h.c.} \qquad \text{and} \qquad \tfrac14 Y^{ABE} Y_{ECD} \left( \phi_A^{\phantom{*}}\, \phi_B^{\phantom{*}} \, \phi_C^* \, \phi_D^*\right).
\end{equation}
As the $\beta$-function for the Yukawa interaction is given by
\begin{equation}
  \beta^{ABC} = \gamma^A_{\phantom{A}D} Y^{DBC} + \gamma^B_{\phantom{A}D} Y^{ADC} + \gamma^C_{\phantom{A}D} Y^{ABD},
\end{equation}
the renormalisation group evolution of the quartic $ \lambda^{AB}_{\phantom{AB}CD} = Y^{ABE} Y_{ECD}$ is constructed via 
\begin{equation}\label{eq:susy-quartic}
\begin{aligned}
    \beta^{AB}_{\phantom{AB}CD} &= \left(\gamma^A_{\phantom{A}F} Y^{FBE} + \gamma^B_{\phantom{F}F} Y^{AFE} \right) Y_{ECD} +  Y^{ABE} \left(Y_{EFD} \,\gamma^F_{\phantom{F}C}  + Y_{ECF} \, \gamma^F_{\phantom{F}D} \right) \\
    &\phantom{=\ } + 2\, Y^{ABF} \gamma^E_{\phantom{F}F}\, Y_{ECD} \,.
\end{aligned}
\end{equation}
Here, $\gamma^A_{\phantom{A}B}$ denotes the chiral superfield anomalous dimensions, with the three-loop part given by \cite{Jack:1996qq,Parkes:1985hh}
\begin{equation}\label{eq:susy_adm}
\begin{aligned}
  (4\pi)^6\gamma^A_{\phantom{B}B} &= Y^{ACD} Y_{DEF} Y^{FGH} Y_{GHI} Y^{EIJ} Y_{BCJ} \\
  &\phantom{=\ } - \tfrac18 \,Y^{ACD} Y_{CEF} Y^{EFG} Y_{DHI} Y^{HIJ} Y_{BGJ} \\
 &\phantom{=\ }  - \tfrac14 \,Y^{ACD} Y_{DEF} Y^{EFG} Y_{GHI} Y^{HIJ} Y_{BCJ}  \\
   &\phantom{=\ }+ \tfrac32 \zeta_3 \, Y^{ACD} Y_{CEF} Y_{DGH} Y^{EGI} Y^{FHJ} Y_{BIJ}\,.
\end{aligned}
\end{equation}
To compare to non-supersymmetric RGEs, the QFT defined by the action
\begin{equation}
  \begin{aligned}
    \mathcal{L} = & \  \partial^\mu \phi^{*}_A\partial_\mu \phi^{\phantom{*}}_A + i\psi_A^\dagger \sigma^\mu \partial_\mu \psi^{\phantom{\dagger}}_A \\
    &-  \tfrac12 \left[ y^{ABC} \left(\phi_A^{\phantom{\intercal}}  \psi_B^\intercal \varepsilon \psi_C^{\phantom{\intercal}} \right) + y^{*ABC} \left(\phi_A^{*}  \psi_B^\dagger \varepsilon \psi_C^{*} \right)\right] - \tfrac14\,\lambda^{AB}_{\phantom{AB}CD} \left(\phi_A^{\phantom{*}}\, \phi_B^{\phantom{*}} \, \phi_C^* \, \phi_D^*\right),
  \end{aligned}
\end{equation}
consisting of Weyl fermions $\psi_A$ and complex scalars $\phi_A$ is analysed. The  $\beta$-function of the quartic $\lambda^{AB}_{\phantom{AB}CD}$ can be computed using eq.\eq{quartic-sum}ff. After insertion of the supersymmetry relations 
\begin{equation}\label{eq:susy-insertion}
  y^{ABC} = Y^{ABC}\,,\qquad
  y^{*ABC} = Y_{ABC}\,,\qquad
  \lambda^{AB}_{\phantom{AB}CD} = Y^{ABE} Y_{ECD}\,,
\end{equation}
the result can be directly compared to \eq{susy-quartic}, constraining unknown coefficients $c_{(n)}$ of tensor structures $(n)$ in \fig{quart1} and \fig{quart2}. 
However, certain features are inaccessible: contractions (5), (18), (19), (23), (24), (27), (30), (33), (37)--(40), (42), (44), (48)--(51), (53), (55) and (56) vanish, since all scalars are complex and the Yukawa interaction reflects the holomorphism of the superpotential. 
Choosing a basis of contractions of $Y^{ABC}$ and $Y_{ABC}$, after the insertion of \eq{susy-insertion}, the extracted relations are listed in \tab{susy-relations}.

\clearpage
\bibliographystyle{JHEP}
\bibliography{ref.bib}

\end{document}